\documentclass[12pt]{article}
\usepackage{epsfig}
\usepackage{subfigure}
\usepackage{amsmath}
\usepackage{amssymb}
\usepackage{latexsym}
\usepackage{hyperref}

\def\Bra#1{\mathinner{\langle{#1}|}}
\def\Ket#1{\mathinner{|{#1}\rangle}}

\def\bra#1{\left<#1\right|}
\def\ket#1{\left|#1\right>}
{\catcode`\|=\active 
  \gdef\Braket#1{\left<\mathcode`\|"8000\let|\bravert {#1}\right>}}
\def\bravert{\egroup\,\vrule\,\bgroup}
\newcommand{\alg}[1]{\mathfrak{#1}}

\newcommand{\su}{\alg{su}}
\newcommand{\Sl}{\alg{sl}}
\newcommand{\so}{\alg{so}}
\newcommand{\tr}{\mathop{\rm tr}}

\newcommand{\be}{\begin{eqnarray}}
\newcommand{\ee}{\end{eqnarray}}
\newcommand{\bea}{\begin{eqnarray}}
\newcommand{\eea}{\end{eqnarray}}
\newcommand{\ben}{\begin{equation}}
\newcommand{\een}{\end{equation}}

\newcommand{\Rhat}{\widehat R}

\newcommand{\nn}{\nonumber}

\numberwithin{equation}{section}

\begin{document}

\begin{titlepage}
\begin{flushright}
CALT-68-2499\\
PUPT-2116\\
hep-th/0405153
\end{flushright}
\vspace{15 mm}
\begin{center}
{\huge Higher Impurity AdS/CFT Correspondence \\
	 in the Near-BMN Limit}
\end{center}
\vspace{12 mm}

\begin{center}
{\large Curtis G.\ Callan, Jr.${}^{a}$,
Tristan McLoughlin${}^{b}$,
Ian Swanson${}^{b}$ }\\
\vspace{3mm}
${}^a$ Joseph Henry Laboratories\\
Princeton University\\
Princeton, New Jersey 08544, USA\\
\vspace{0.5 cm}
${}^b$ California Institute of Technology\\
Pasadena, CA 91125, USA 
\end{center}
\vspace{5 mm}
\begin{center}
{\large Abstract}
\end{center}
\noindent

The pp-wave/BMN limit of the AdS/CFT correspondence has exposed the
Maldacena conjecture to a new regimen of direct tests.  In one line
of pursuit, finite-radius curvature corrections to the Penrose limit
(which appear in inverse powers of the string angular momentum $J$)
have been found to induce a complicated system of interaction perturbations to 
string theory on the pp-wave; these have been successfully matched
to corresponding corrections to the BMN dimensions of ${\cal N}=4$
super Yang-Mills (SYM) operators to two loops in the 't Hooft coupling
$\lambda$.  This result is tempered by a well-established 
breakdown in the correspondence at three loops.   
Notwithstanding the third-order mismatch, we proceed with this line of 
investigation by subjecting the string and gauge theories 
to new and significantly more rigorous tests.  Specifically, we extend 
our earlier results at $O(1/J)$ in the curvature expansion
to include string states and SYM operators 
with three worldsheet or $R$-charge impurities.  In accordance with
the two-impurity problem, we find a perfect and intricate agreement 
between both sides of the correspondence to two-loop order in $\lambda$ and, 
once again, the string and gauge theory predictions fail to agree
at third order.  

\vspace{1cm}
\begin{flushleft}
\today
\end{flushleft}
\end{titlepage}
\newpage
\section{Introduction}

Recent explorations of the AdS/CFT correspondence have focused on comparing 
the spectrum of the type IIB superstring boosted to a 
near-lightcone (large angular momentum $J$) trajectory on $AdS_5\times S^5$ 
with the anomalous dimensions of large $R$-charge operators 
in ${\cal N}=4$ supersymmetric Yang-Mills (SYM) theory.  This line of investigation was 
launched on the string side by Metsaev \cite{Metsaev:2001bj,Metsaev:1998it}, 
who showed that the string 
worldsheet dynamics can be reduced to a system of free massive fields by taking
a Penrose limit of the target geometry that reduces $AdS_5\times S^5$
to a pp-wave.  On the gauge theory side, Berenstein, Maldacena and Nastase 
\cite{Berenstein:2002jq} 
identified the relevant class of gauge theory operators (commonly referred to as 
BMN operators) whose dimensions can be calculated perturbatively and matched 
to the string theory energy spectrum on the pp-wave. 

Attempts to push the original results further have gone in
two independent directions.  In the gauge theory, the calculation of
anomalous dimensions of BMN operators has been greatly simplified by Minahan and Zarembo's
discovery that the problem can be mapped to that of computing the energies of
certain integrable spin chains \cite{Minahan:2002ve}.  Based on this development,
calculations in certain sectors of the theory
have been carried out to three loops in the 't~Hooft coupling $\lambda$
\cite{Beisert:2003tq,Beisert:2003ys}.\footnote{We note that the conjectural
three-loop computation of \cite{Beisert:2003tq} was solidified by field
theoretic methods in \cite{Beisert:2003ys}.} 
At the same time, the quantization of 
the Green-Schwarz string in the $AdS_5\times S^5$ background has developed 
far enough to enable perturbative computations of the effect of worldsheet 
interactions on the spectrum of the string when it is boosted to large, but finite, 
angular momentum $J$ \cite{Parnachev:2002kk,Callan:2003xr,STRpaper}. 
These two approaches lead to different 
expansions of operator anomalous dimensions (or string eigenenergies): on the 
gauge theory side, one naturally has an expansion in the coupling constant 
$\lambda$ which is typically exact in $R$-charge; on the 
string theory side one has an expansion in inverse powers of angular momentum $J$
(the dual of gauge theory $R$-charge) which is exact in $\lambda$. 

The expansion on the string side is difficult and has so far been carried out
to ${O}(1/J)$ for `two-impurity' states (i.e.~states with two 
string oscillators excited). The resulting functions of the loop expansion parameter 
$\lambda$ can be compared with the large $R$-charge expansion of two-impurity BMN operators 
in the gauge theory to provide new and stringent tests of the AdS/CFT 
correspondence.  As mentioned above, recently developed gauge theory technology
has made it possible to compute anomalous dimensions of certain two-impurity
BMN operators out to three-loop order. The agreement between dual quantities 
is perfect out to two-loop order but, surprisingly, seems to break down at three loops 
\cite{Callan:2003xr,STRpaper}.  
Exactly what this means for the AdS/CFT correspondence is not yet clear but, 
given the circumstances, it seems appropriate to at least look for further 
data on the disagreement in the hope of finding some instructive systematics.  
The subject of this paper is to pursue one possible line of attack in which 
we extend the calculations described above to higher-impurity string states 
and gauge theory operators.  The extension of our two-impurity results to 
higher impurities is not a straightforward matter on either side of the 
correspondence and gets more complex as the number of impurities increases. 
We focus here on the three-impurity case, where we obtain results which validate 
our methods for quantizing the Green-Schwarz superstring; the agreement with gauge theory
at one and two loops is impressive, though we also confirm the previously observed
breakdown of agreement at three-loop order.

In section 2, we give a brief review of
string quantization on $AdS_5\times S^5$. Since we
use exactly the same methods and notation as in our recent treatment
of the two-impurity problem, we can refer the reader to our long paper
on that topic for a more detailed exposition of our methods \cite{STRpaper}. 
In sections 3 and 4 we present the details of the diagonalization of
the perturbing string worldsheet Hamiltonian on degenerate subspaces
of three-impurity states.  We give a
compressed discussion of general strategy, concentrating on the
aspects of the problem which are new to the three-impurity case.
An interesting new element is that the non-interacting degenerate 
subspace breaks up into several different supersymmetry multiplets 
so that the detailed accounting of multiplicities and irrep
decomposition amounts to a stringent test that the quantization
has maintained the correct nonlinearly realized superconformal
symmetries.  Section 5 is devoted to the comparison of the string
theory spectrum with gauge theory anomalous dimensions. Since
gauge theory results have not been obtained elsewhere for higher-impurity
operators to the order we need, we generate our own data by doing 
numerical analysis, along the lines of \cite{Beisert:2003ea}, 
of the various higher-loop spin chains onto which
the gauge theory anomalous dimension problem has been mapped.
This is an interesting problem in its own right and we give only
a brief description of our methods, referring the reader to
a separate methods publication \cite{spinchain} for details. We are able
to take scaling limits of the numerical spin chain analyses
that allow us to make clean contact with the string theory 
results.  We find perfect agreement through two-loop order and,
once again, breakdown at three loops. Section 6 is devoted
to discussion and conclusions.

Overall, the three-impurity regime of the string theory
offers a much more stringent test of the duality away from
the full plane-wave limit. While we are unable to offer a solution
to the disagreement with gauge theory at three loops, we can 
confirm that that the complicated interacting worldsheet theory
at $O(1/\Rhat^2)$ in the curvature expansion is properly quantized 
and correct to two loops in $\lambda$.  

\section{String quantization on $AdS_5\times S^5$: brief review}

A standard presentation of the $AdS_5\times S^5$ metric is
\begin{equation}
\label{adsmetricPre}
ds^2 = \Rhat^2 ( - {\rm cosh}^2 \rho\ dt^2 + d \rho^2 + {\rm sinh}^2
\rho\ d \Omega_3^2 + {\rm cos}^2  \theta\ d \phi^2 +  d \theta^2 +
{\rm sin}^2 \theta\ d \tilde\Omega_3^2)~,
\end{equation}
where $\Rhat$ is the radius of each subspace, and $d \Omega^2$,
$d \tilde\Omega_3^2$ are separate three-spheres.  The spacetime radius is 
related to the coupling of the dual $SU(N_c)$ gauge theory by 
$\Rhat^4 = \lambda ({\alpha'})^2 =  g_{YM}^2 N_c ({\alpha'})^2$.
The reparameterization 
\begin{eqnarray}
    \cosh\rho  =  \frac{1+z^2/4}{1-z^2/4} \qquad
    \cos\theta  =  \frac{1-y^2/4}{1+ y^2/4}\ 
\end{eqnarray}
simplifies the spin connection and is particularly convenient for constructing
the superstring action in the coset formalism \cite{Callan:2003xr,STRpaper}.
In these coordinates, the target-space metric takes the form
\begin{equation}
\label{metric}
ds^2  =  \Rhat^2
\biggl[ -\left({1+ \frac{1}{4}z^2\over 1-\frac{1}{4}z^2}\right)^2dt^2
        +\left({1-\frac{1}{4}y^2\over 1+\frac{1}{4}y^2}\right)^2d\phi^2
    + \frac{d z_k dz_k}{(1-\frac{1}{4}z^2)^{2}}
    + \frac{dy_{k'} dy_{k'}}{(1+\frac{1}{4}y^2)^{2}} \biggr]~.
\end{equation}
The transverse $SO(8)$ breaks into $SO(4)\times SO(4)$, which is
spanned by $z^2 = z_k z^k$ with $k=1,\dots,4$, and $y^2 = y_{k'} y^{k'}$ 
with $k'=5,\dots,8$.
We use lightcone coordinates
\be
\label{newcoords}
t  =  x^+\, , \qquad
\phi  =  x^+ +{x^-}/{\Rhat^2}\ ,
\ee
leading to conjugate lightcone momenta
\be
\label{ltconmom}
-p_+  =  \Delta - J~, \qquad
-p_-  =  i\partial_{x^-} = \frac{i}{\Rhat^2}\partial_\phi = -\frac{J}{\Rhat^2}\ ,
\ee
where $J$ is the conserved angular momentum conjugate to $\phi$ and $\Delta$ is
the conserved energy conjugate to $t$ (which will eventually be identified with gauge theory
operator dimensions). The technical benefits of this lightcone coordinate choice 
are explained in \cite{Callan:2003xr,STRpaper}). When quantizing the string in the
BMN limit, $p_-$ is held fixed while $J$ (the angular momentum of the string)
and $\Rhat$ (the scale of the geometry) are both taken to be large. This is possible
because $p_- \Rhat^2 = J$ (\ref{ltconmom}) and we can regard this as either
a large-$J$ or large-$\Rhat$ limit; we will pass freely between the two.

The large-$\Rhat$ expansion of (\ref{metric}) is
\be
\label{Rhatexp}
ds^2 & \approx & 2dx^+ dx^- - (x^A)^2 (dx^+)^2 + (dx^A)^2 
\nonumber \\
& & 	+ \frac{1}{\Rhat^2}\left[ 
	-2y^2 dx^+ dx^- + \frac{1}{2}(y^4-z^4) (dx^+)^2 + (dx^-)^2
	+\frac{1}{2}z^2 dz^2 - \frac{1}{2}y^2 dy^2 \right] 
\nonumber \\
\label{expmetric2}
& & 	+ {\cal O}\left({1}/{\Rhat^4}\right)\ .
\ee
When this metric is used to construct the worldsheet string action, the
$O(1/\Rhat^0)$ terms lead to a quadratic (free) theory of worldsheet fields
and the $O(1/\Rhat^2)$ terms lead to quartic interaction terms. We will give
a perturbative treatment of the effect of these interactions on the
energy spectrum of the leading free theory.

The Green-Schwarz action of type IIB superstring theory on
$AdS_5\times S^5$ can be expressed as a non-linear sigma model
constructed from supersymmetric Cartan one-forms $L_a^\mu$
(the index $a$ is a worldsheet index here) on the coset manifold
$G/H = [SO(4,2)\times SO(6)]/[SO(4,1)\times SO(5)]$ \cite{Metsaev:1998it}.
The superconformal algebra of the coset manifold takes the generic form
\begin{eqnarray}
\label{genalgebra}
\left[ B_\mu, B_\nu \right]  =  f_{\mu\nu}^\rho B_\rho \qquad
\left[ F_\alpha, B_\nu \right]  =  f_{\alpha \nu}^\beta F_\beta \qquad
\{ F_\alpha, F_\beta \}  =  f_{\alpha\beta}^\mu B_\mu\ ,
\end{eqnarray}
where $B_\mu$ ($F_\alpha$) represent bosonic (fermionic) generators.  
The Cartan forms $L^\mu$ and superconnections $L^\alpha$
are determined by the structure constants
$f_{\alpha \mu}^J$ and $f_{\alpha\beta}^\mu$ according to
\begin{eqnarray}
\label{kallosh1}
L_{at}^\alpha & = &
	\left( \frac{\sinh t{\cal M}}{\cal M} \right)^\alpha_\beta 
	({\cal D}_a \theta)^\beta
 \\
\label{kallosh2}
L_{at}^\mu & = & e^\mu_{\phantom{\mu}\nu}\partial_a x^\nu + 2\theta^\alpha f_{\alpha\beta}^\mu
	\left( \frac{\sinh^2 (t {\cal M}/2)}{{\cal M}^2}\right)^\beta_\gamma 
	({\cal D}_a \theta)^\gamma
 \\
({\cal M}^2)^\alpha_\beta & = & -\theta^\gamma f^\alpha_{\gamma\mu} 
	\theta^\delta f^\mu_{\delta\beta}\ ,
\end{eqnarray}
with $L_a^\mu = {L_{at}^\mu}|_{t=1}$.
In $AdS_5 \times S^5$, the Lagrangian takes the form
\begin{eqnarray}
\label{lagrangiank}
{\cal L}_{\rm Kin} & = & -\frac{1}{2} h^{ab} L_a^\mu L_b^\mu  
\\
\label{lagrangianwz}
{\cal L}_{\rm WZ} & = & -2i\epsilon^{ab} \int_0^1 dt\, L_{at}^\mu s^{IJ}
    \bar\theta^I \Gamma^\mu L_{bt}^J\ ,
\end{eqnarray}  
where $\Gamma^\mu$ are $SO(9,1)$ Dirac gamma matrices,
$\eta_{\mu\nu}$ is the $SO(9,1)$ Minkowski metric and $s^{IJ}={\rm diag}(1,-1)$.
The worldsheet fermi fields $\theta^I$ ($I,J=1,2$)
are two $SO(9,1)$ Majorana--Weyl spinors of the same chirality
$\Gamma_{11}\theta^I=\theta^I$. The $\kappa$-symmetry condition
$\Gamma_0\Gamma_9\theta^I=\theta^I$ sets half the components of $\theta^I$ 
to zero. In constructing the Hamiltonian, it will be convenient to 
define $\psi = \sqrt{2}(\theta^1 + i\theta^2)$:  
$\psi_\alpha$ is an eight-component complex spinor built out of the 16 
components of $\theta^I$ that survive the gauge fixing. It will be useful
to split components of $\psi$ according to their transformation under
the parity operator $\Pi \equiv \gamma^1 \bar\gamma^2 \gamma^3 \bar\gamma^4$,
where $\gamma^a,\bar\gamma^a$ are $8\times 8$ $SO(8)$ gamma matrices.

Ultimately, we want to expand the GS worldsheet Hamiltonian in powers of 
$1/\Rhat$, along the lines of (\ref{Rhatexp}), in order to develop a 
perturbation theory of the string spectrum. To do this, we expand all the
elements of the GS action (metric, spin connections, one-forms, etc.) in 
powers of $1/\Rhat$ and write the lightcone coordinates $x^+$ and $x^-$ 
in terms of the physical transverse fields (by solving the $x^-$ 
equations of motion and the covariant gauge constraints order-by-order). 
Explicit quantization rules are obtained by rewriting everything in terms 
of unconstrained canonical variables after removing the second-class
constraints to which some of the fermionic canonical variables are subject.
This awkward and rather complicated procedure is described in detail in 
\cite{Callan:2003xr,STRpaper}. The final result for the Hamiltonian 
density, correct to $O(1/J)$, is
\begin{eqnarray}
\label{GSHamFinal}
{H}={H}_{\rm pp}+{H}_{\rm int}\ , \qquad
{H}_{\rm int}={H}_{\rm BB}+{H}_{\rm FF}+{H}_{\rm BF}~,
\end{eqnarray}
where
\begin{eqnarray}
{H}_{\rm pp} & = &
    \frac{1}{2}\left[(x^A)^2 + (p_A)^2 + ({x'}^A)^2\right]
    + \frac{i}{2}\left[
     \psi\psi' -  \rho \rho' + 2\rho\Pi\psi \right]\ ,
\end{eqnarray}
\begin{eqnarray}
\label{HBBfinal}
{H}_{\rm BB} & = & \frac{1}{\Rhat^2}\biggl\{
    \frac{1}{4}\left[ z^2\left( p_{y}^2 + {y'}^2 + 2{z'}^2 \right)
    -y^2\left( p_z^2 + {z'}^2 + 2{y'}^2\right)\right]
    + \frac{1}{8}\left[ (x^A)^2 \right]^2
\nn\\
& &     - \frac{1}{8}\left[
    \left[ (p_A)^2\right]^2 + 2(p_A)^2({x'}^A)^2
    + \left[ ({x'}^A)^2\right]^2 \right]
     + \frac{1}{2}\left({x'}^A p_A\right)^2
    \biggr\}\ ,
\end{eqnarray}
\begin{eqnarray}
{H}_{\rm FF} & = &
        -\frac{1}{4 \Rhat^2}\biggl\{
  \left[ (\psi'\psi) + (\rho\rho')\right](\rho\Pi\psi)
        -\frac{1}{2}(\psi'\psi)^2 - \frac{1}{2}(\rho'\rho)^2
        + (\psi'\psi)(\rho'\rho)
\nn\\
& &     + (\rho\psi')(\rho'\psi)
    -\frac{1}{2}\left[ (\psi\rho')(\psi\rho') + (\psi'\rho)^2\right]
        + \biggl[
    \frac{1}{12}(\psi\gamma^{jk}\rho)(\rho\gamma^{jk}\Pi\rho')
\nn\\ & &     -\frac{1}{48}
        \left(\psi\gamma^{jk}\psi - \rho\gamma^{jk}\rho\right)
        \left(\rho'\gamma^{jk}\Pi\psi - \rho\gamma^{jk}\Pi\psi'\right)
        - (j,k \rightleftharpoons j',k') \biggr] \biggr\}\ ,
\end{eqnarray}
\begin{eqnarray}
\label{HBFfinal}
{H}_{\rm BF} & = &
     \frac{1}{\Rhat^2}\biggl\{ -\frac{i}{4}\left[(p_A)^2+({x'}^A)^2
    + (y^2 - z^2)\right]\left(\psi\psi'- \rho\rho'\right)
\nn\\ & &     -\frac{1}{2}(p_A{x'}^A)(\rho\psi' + \psi\rho' )
        -\frac{i}{2}\left( p_k^2 + {y'}^2 -  z^2 \right)\rho\Pi\psi
\nn\\
& &     +\frac{i}{4}(z'_j z_k)\left(\psi\gamma^{jk}\psi -
    \rho\gamma^{jk}\rho\right)
        -\frac{i}{4}(y'_{j'} y_{k'})\left(\psi\gamma^{j'k'}\psi -
   \rho\gamma^{j'k'}\rho\right)
\nn\\
& &     -\frac{i}{8}(z'_k y_{k'} + z_k y'_{k'})
       \left(\psi\gamma^{kk'}\psi - \rho\gamma^{kk'}\rho\right)
        +\frac{1}{4{}}(p_k y_{k'} +  z_k p_{k'} )\psi\gamma^{kk'}\rho
\nn\\ & &     +\frac{1}{4{}}(p_j z'_k)\left(\psi\gamma^{jk}\Pi\psi
                + \rho\gamma^{jk}\Pi\rho\right)
        -\frac{1}{4{}}(p_{j'} y'_{k'})\left(\psi\gamma^{j'k'}\Pi\psi
                + \rho\gamma^{j'k'}\Pi\rho\right)
\nn\\ & &     
 -\frac{i}{2}(p_kp_{k'} - z'_k y'_{k'})\psi\gamma^{kk'}\Pi\rho  \biggr\}\ .
\end{eqnarray}

The canonical commutation relations of $p^A,\ x^A$ and $\psi,\ \rho$ mean that
they can be expanded in bosonic creation operators $a_n^{A\dag}$,  
fermionic creation operators $b_n^{\alpha\dag}$ and their annihilation
operator counterparts (where $n$ is a mode index). 
Upper-case Latin letters $A,B,C,\dots \in 1,\dots,8 $ indicate vectors of the
transverse $SO(8)$: these are divided into the two $SO(4)$
subgroups which descend from the $AdS_5$ and $S^5$ subspaces. 
Lower-case letters $a,b,c,\dots \in 1,\dots,4$ label vectors
in $SO(4)_{AdS}$, while $a',b',c',\dots \in 5,\dots,8$ 
indicate vectors in $SO(4)_{S^5}$.  Spinors of $SO(9,1)$ are
labeled by $\alpha,\beta,\gamma,\dots \in 1,\dots,8$. All told, 
there are 16 oscillators per mode $n$. The Fock space of physical string 
states is generated by acting on the ground state $\Ket{J}$ (which 
carries the total angular momentum of the string state) with arbitrary
combinations of the above creation operators. Matrix elements 
of the Hamiltonian are computed by expanding each of the fields and 
conjugate momenta of ${H}$ in the mode creation and annihilation 
operators and acting on the Fock space in the obvious way.
 
The parameter equivalences between the string and gauge theories
are determined by a modified AdS/CFT dictionary which emerges in the 
pp-wave/BMN limit. The $R$-charge is equated on the string side with 
$p_-\Rhat^2 = J$, and the large $R$-charge limit corresponds to
taking $J = p_-\Rhat^2 \to \infty$ and $N_c \to \infty$, keeping the 
ratio $N_c/J^2$ fixed.  The gauge theory coupling $\lambda = g_{YM}^2 N_c$ 
is then replaced on the string side with the so-called modified 
't~Hooft coupling $\lambda' = g_{YM}^2 N_c/J^2$.

At leading order in $J^{-1}$, ${H}$ reduces to ${H}_{\rm pp}$
and the string theory reduces to a theory of eight free massive bosons 
and fermions.
Altogether, the sixteen oscillators for each mode $n$ contribute 
$\omega_n = \sqrt{1+n^2\lambda'}$ to the energy of the string 
and produce a highly degenerate spectrum. 
For example, the 256 `two-impurity' states spanned by
\be
a_n^{A\dag} a_{-n}^{B\dag}\Ket{J} & \quad &
b_n^{\alpha\dag} b_{-n}^{\beta\dag}\Ket{J}\ \qquad {\rm (spacetime\ bosons)}  \nn\\
a_n^{A\dag} b_{-n}^{\beta\dag}\Ket{J} & \quad &  
b_n^{\alpha\dag} a_{-n}^{B\dag}\Ket{J}\ \qquad {\rm (spacetime\ fermions)}
\ee
all have the same energy $E=2\sqrt{1+n^2\lambda'}$ in the large-$J$ limit
(note that the mode indices of the oscillators must sum to zero due
to the level-matching constraint). This degeneracy is broken at the
first non-leading order in $J^{-1}$ by the action of the perturbing
Hamiltonian $H_{\rm int} = {H}_{\rm BB}+{H}_{\rm FF}+{H}_{\rm BF}$. 

To find the spectrum, one must diagonalize the 
$256\times 256$ perturbation matrix schematically represented in 
table~\ref{blockform}.
\begin{table}[ht!]
\begin{eqnarray}
\begin{array}{|c|cccc|}
\hline
 ({H})_{\rm int} & a^{A\dagger}_n a^{B\dagger}_{-n} \ket{J} &
        b^{\alpha\dagger}_n b^{\beta\dagger}_{-n}\ket{J} &
        a^{A\dagger}_n b^{\alpha\dagger}_{-n} \ket{J} &
        a^{A\dagger}_{-n} b^{\alpha\dagger}_{n} \ket{J} \\
        \hline
\bra{J} a^{A}_n a^{B}_{-n} & { H}_{\rm BB} & { H}_{\rm BF} &0&0 \\
\bra{J} b^{\alpha}_n b^{\beta}_{-n} & { H}_{\rm BF} & { H}_{\rm
FF}&0&0\\ \bra{J} a^{A}_n b^{\alpha}_{-n} &0&0& { H}_{\rm BF} & {
H}_{\rm BF} \\ \bra{J} a^{A}_{-n} b^{\alpha}_n & 0 & 0 & { H}_{\rm
BF} & { H}_{\rm BF}\\
\hline
\end{array} \nonumber
\end{eqnarray}
\caption{Complete Hamiltonian in the space of two-impurity string states}
\label{blockform}
\end{table}
The matrix of course block-diagonalizes on spacetime bosons and spacetime 
fermions, so the problem is actually $128\times 128$. The matrix elements 
of the non-vanishing sub-blocks are computed by expanding each of the
fields (and conjugate momenta) in ${H}_{\rm BB},\ {H}_{\rm FF}$ and ${H}_{\rm BF}$ 
in mode creation and annihilation operators and evaluating the indicated
Fock space matrix elements. Finding the perturbed spectrum is then
a matter of diagonalizing the explicit matrix constructed in this
fashion.  The calculation is algebraically tedious and requires the
use of symbolic manipulation programs, but the end results are quite
simple and verify, as previously described, a perfect match of the string
spectrum to the gauge theory results out to two-loop order. 

Our purpose in this paper is to work out the analogous results for
higher-impurity states on both sides of the correspondence. 
In fact, we will limit our detailed considerations to the three-impurity problem, 
since it presents many interesting
complications as compared to the two-impurity case, and it is not clear
that any useful illumination will come from studying yet higher-impurity
states.

\section{Three-impurity string spectrum: leading order in $\lambda'$}

The three-impurity Fock space block-diagonalizes into separate spacetime 
fermion and spacetime boson sectors. The bosonic sector contains states 
that are purely bosonic (composed of three bosonic string oscillators) and 
states with bi-fermionic components:
\be
a_q^{A\dag} a_r^{B\dag} a_s^{C\dag}\ket{J}
	\qquad a_q^{A\dag} b_r^{\alpha\dag} b_s^{\beta\dag}\ket{J}\ .
\label{state1}
\ee
Pure boson states are mixed by the bosonic sector of the Hamiltonian
$H_{\rm BB}$, while states with bi-fermionic excitations are mixed both by the purely
fermionic Hamiltonian $H_{\rm FF}$ and the bose-fermi sector $H_{\rm BF}$.
The sector of spacetime fermion states is composed of purely fermionic excitations 
and mixed states containing two bosonic oscillators:
\be
b_q^{\alpha\dag} b_r^{\beta\dag} b_s^{\gamma\dag}\ket{J}
	\qquad a_q^{A\dag} a_r^{B\dag} b_s^{\alpha\dag}\ket{J}\ .
\label{state2}
\ee
Pure fermion states are acted on  by $H_{\rm FF}$, and mixed states with 
bosonic excitations are acted on by $H_{\rm BB}$ and $H_{\rm BF}$. This block 
diagonalization of the perturbing Hamiltonian is displayed schematically 
in table~\ref{blockform3}.
\begin{table}[ht!]
\begin{eqnarray}
\begin{array}{|c|cccc|}
\hline
 ({H})_{int} & a^{A\dagger} a^{B\dagger} a^{C\dagger} \ket{J} &
        a^{A\dag} b^{\alpha\dagger} b^{\beta\dagger} \ket{J} &
        b^{\alpha\dagger} b^{\beta\dagger} b^{\gamma\dag} \ket{J} &
        a^{A\dag} a^{B\dag} b^{\alpha\dag} \ket{J} \\
        \hline
\bra{J} a^{A} a^{B} a^C & { H}_{BB} & { H}_{BF} &0&0 \\
\bra{J} a^A b^{\alpha} b^{\beta} & { H}_{BF} & H_{FF}+ { H}_{BF}&0&0\\ 
\bra{J} b^{\alpha}  b^{\beta}  b^\gamma  &0&0& { H}_{FF} & {H}_{BF} \\ 
\bra{J} a^{A}  a^{B}  b^{\alpha}  & 0 & 0 & { H}_{BF} & { H}_{BB} + H_{BF} \\
\hline
\end{array} \nonumber
\end{eqnarray}
\caption{Three-impurity string states}
\label{blockform3}
\end{table}

The three-impurity string states are subject to the usual level-matching 
condition on the mode indices: $q+r+s=0$. There are two generically 
different solutions of this constraint: all mode indices different ($q\neq r\neq s$)
and two indices equal (eg.~$q=r=n,\ s=-2n$).
In the inequivalent index case, there are $16^3=4,096$ degenerate states
arising from different choices of spacetime labels on the mode creation operators.
In the case of two equivalent indices, the dimension of the degenerate subspace
is half as large (there are fewer permutations on mode indices that generate 
linearly independent states). The two types of basis break up into irreducible 
representations of $PSU(2,2|4)$ in different ways and must be studied separately.

As in the two-impurity case, the problem of diagonalizing the perturbation
simplifies enormously when the matrix elements are expanded to leading order
in $\lambda'$.  We will take this approach here
to obtain an overview of how degeneracies are lifted by the interaction.
The generalization of the results to all loop orders in $\lambda'$ (but still 
to first non-leading order in $1/J$) will be presented in the next section.
We use the same methods, conventions and notations as in our recent detailed
study of the two-impurity problem \cite{STRpaper} (especially in Sec.~6 of 
that paper). It is once again the case that in the one-loop approximation, 
projection onto invariant subspaces under the manifest global $SO(4)\times SO(4)$ 
symmetry often diagonalizes the Hamiltonian directly (and at worst reduces 
it to a low-dimensional matrix).  Symbolic manipulation programs
were used to organize the complicated algebra and to perform explicit
projections onto invariant subspaces. 

\subsection{Matrix evaluation: inequivalent mode indices $(q\neq r\neq s)$}

In the sector of spacetime bosons, the subspace
of purely bosonic states $a_q^{A\dag} a_r^{B\dag} a_s^{C\dag}\ket{J}$ 
is 512-dimensional.  When each of the three mode indices $(q,r,s)$ are 
different, states with bi-fermionic excitations 
$a_q^{A\dag} b_r^{\alpha\dag} b_s^{\beta\dag}\ket{J}$
are inequivalent under permutation of the mode indices, and form a 
1,536-dimensional subsector. The entire bosonic sector of the three-impurity 
state space therefore contains 2,048 linearly independent states.  
The fermionic sector decomposes in a similar manner:
the subsector of purely fermionic states 
$b_q^{\alpha\dag} b_r^{\beta\dag} b_s^{\gamma\dag}\ket{J}$  
is 512-dimensional; fermionic states containing two bosonic excitations 
$a_q^{A\dag} a_r^{B\dag} b_s^{\alpha\dag}\ket{J}$ 
are inequivalent under permutation of the mode indices, and comprise
an additional 1,536-dimensional subsector. Adding this 2,048-dimensional 
fermion sector brings the dimensionality of the entire state space to 4,096.  

Our first task is to evaluate the interaction Hamiltonian matrix. 
The matrix elements needed to fill out the spacetime boson sector 
are listed in table~\ref{boseblock3}. To evaluate the entries, 
we express the Hamiltonians (\ref{HBBfinal}-\ref{HBFfinal}) in terms of mode creation and 
annihilation operators, expand the result in powers of $\lambda'$ and then 
compute the indicated matrix elements between three-impurity Fock space states.
We collect below all the relevant results of this exercise.
\begin{table}[ht!]
\begin{eqnarray}
\begin{array}{|c|cccc|}
\hline
 {H}_{\rm int} & 
	a_s^{D\dag}  a^{E\dagger}_r  a^{F\dagger}_{q}\ket{J} &
	a_s^{D\dag}  b^{\gamma\dagger}_r  b^{\delta\dagger}_{q}\ket{J} &
        a_r^{D\dag}  b^{\gamma\dagger}_q  b^{\delta\dagger}_{s}\ket{J} &
	a_r^{D\dag}  b^{\gamma\dagger}_s  b^{\delta\dagger}_{q}\ket{J}
\\   \hline
\bra{J} a^{A}_q  a^{B}_r  a^{C}_{s}      & H_{\rm BB}    & H_{\rm BF}
						 & H_{\rm BF} & H_{\rm BF} \\
\bra{J} a^{A}_q  b^{\alpha}_r  b^{\beta}_{s}& H_{\rm BF} & H_{\rm FF}+H_{\rm BF} 
						& H_{\rm BF} & H_{\rm BF} \\
\bra{J} a^{A}_s  b^{\alpha}_q  b^{\beta}_{r}& H_{\rm BF} & H_{\rm BF} 
					& H_{\rm FF}+H_{\rm BF} & H_{\rm BF} \\ 
\bra{J} a^{A}_r  b^{\alpha}_s  b^{\beta}_{q}& H_{\rm BF} & H_{\rm BF} 
					& H_{\rm BF} & H_{\rm FF}+ {H}_{\rm BF}\\ 
\hline
\end{array} \nonumber
\end{eqnarray}
\caption{Interaction Hamiltonian on spacetime boson
three-impurity string states $(q\neq r\neq s)$}
\label{boseblock3}
\end{table}

We will use an obvious $(m,n)$ matrix notation to distinguish the different entries in
table~\ref{boseblock3}. The purely bosonic, 512-dimensional $(1,1)$ block has
the explicit form
\be
\Braket{J| a_q^{A} a_r^{B} a_s^{C} (H_{\rm BB})a_s^{D\dag} a_r^{E\dag} a_q^{F\dag}|J}
	& = & 
	\frac{\lambda'}{J}
	\delta^{AF}\delta^{BE}\delta^{CD}
	\left( rs+q(r+s)-q^2-r^2-s^2\right)
\nn\\	+ \frac{\lambda'}{2J}
	\biggl\{
	\delta^{AF}\Bigl[
	(r^2+s^2)
	\bigl(
	\delta^{cd}\delta^{be}&&\kern-25pt  
	-  \delta^{c'd'}\delta^{b'e'}
	\bigr)
	+ (s^2-r^2)
	\bigl(
	\delta^{be}\delta^{c'd'}  
	- \delta^{cd}\delta^{b'e'}
	\bigr)
\nn\\	+2rs\bigl(
	\delta^{bd}\delta^{ce} 
	- \delta^{bc}\delta^{de} &&\kern-25pt 
	- \delta^{b'd'}\delta^{c'e'} 
	+ \delta^{b'c'}\delta^{d'e'}
	\bigr)
	\Bigr]
	+  
	\bigl( 
	r\rightleftharpoons q,\ 
	F\rightleftharpoons E,\ 
	A\rightleftharpoons B
	 \bigr)
\nn\\
&&\kern+65pt	+  \bigl( 
	s\rightleftharpoons q,\ 
	F\rightleftharpoons D,\ 
	A\rightleftharpoons C
	 \bigr)\biggr\}\ .
\label{bosons}
\ee
The off-diagonal entries that mix purely bosonic states
$a_q^{A\dag}a_r^{B\dag}a_s^{C\dag}\Ket{J}$ with states containing bi-fermions
$a_q^{A\dag}b_r^{\alpha\dag}b_s^{\beta\dag}\Ket{J}$ are given by a separate set of 
512-dimensional matrices. The $(1,2)$ block in table~\ref{boseblock3}, for 
example, yields
\be
\Braket{J| a_q^{A} a_r^{B} a_s^{C} (H_{\rm BF}) 
	a_s^{D\dag} b_r^{\alpha\dag} b_q^{\beta\dag}|J}
	& = & 
	\frac{\lambda'}{2J}\delta^{CD} qr
	\biggl\{
	\left(\gamma^{ab'}\right)^{\alpha\beta} - 
	\left(\gamma^{a'b}\right)^{\alpha\beta}
	\biggr\}\ , 
\label{12block}
\ee
where the index $a\ (a')$ symbolizes the value of the vector index $A$, 
provided it is in the first (second) $SO(4)$.
There are six blocks in this subsector, each given by 
a simple permutation of the mode indices $(q,r,s)$ in eqn.~(\ref{12block}).  
In table~\ref{boseblock3}, these matrices occupy the $(1,2)$, 
$(1,3)$ and $(1,4)$ blocks, along with their transposes in the
$(2,1)$, $(3,1)$ and $(4,1)$ entries.  

The pure-fermion sector of the Hamiltonian, $H_{\rm FF}$, has non-vanishing
matrix elements between states containing bi-fermionic excitations.
The $H_{\rm FF}$ contribution to the $(2,2)$ block, for example, is given by
\be
\Braket{J| b_q^{\alpha} b_r^{\beta} a_s^{A} (H_{\rm FF}) 
	a_s^{B\dag} b_r^{\gamma\dag} b_q^{\delta\dag}|J}
	& = & 
	-\frac{\lambda'}{2J}(q-r)^2\delta^{AB}
	\delta^{\alpha\delta}\delta^{\gamma\beta}
\nn\\
	+\frac{\lambda'}{24 J}\delta^{AB} qr\biggl\{
	\bigl(\gamma^{ij}\bigr)^{\alpha\gamma} &&\kern-25pt
	\bigl(\gamma^{ij}\bigr)^{\beta\delta}
	- \bigl(\gamma^{ij}\bigr)^{\alpha\beta}
	\bigl(\gamma^{ij}\bigr)^{\gamma\delta}
	- \bigl(\gamma^{ij}\bigr)^{\alpha\delta}
	\bigl(\gamma^{ij}\bigr)^{\beta\gamma}
\nn\\
	-\bigl(\gamma^{i'j'}\bigr)^{\alpha\gamma} 
	\bigl(\gamma^{i'j'}\bigr)^{\beta\delta} &&\kern-25pt
	+ \bigl(\gamma^{i'j'}\bigr)^{\alpha\beta}
	\bigl(\gamma^{i'j'}\bigr)^{\gamma\delta}
	+ \bigl(\gamma^{i'j'}\bigr)^{\alpha\delta}
	\bigl(\gamma^{i'j'}\bigr)^{\beta\gamma}
	\biggr\}\ .
\label{HFF22}
\ee
A similar contribution, related to this one by simple permutations of
the mode indices $(q,r,s)$, appears in the diagonal blocks
$(3,3)$ and $(4,4)$ as well.

The bose-fermi mixing Hamiltonian $H_{\rm BF}$ makes the following contribution 
to the lower diagonal blocks $(2,2)$, $(3,3)$ and $(4,4)$ in 
table~\ref{boseblock3}:  
\be
\Braket{J| b_q^{\alpha} b_r^{\beta} a_s^{A} (H_{\rm BF}) 
	a_s^{B\dag} b_r^{\gamma\dag} b_q^{\delta\dag}|J}
	& = & 
	\frac{\lambda'}{2J}\biggl\{
	2s(q+r-s)\delta^{ab}\delta^{\alpha\delta}\delta^{\beta\gamma}
	- rs 
	\Bigl[
	\bigl(\gamma^{ab}\bigr)^{\beta\gamma}
	-\bigl(\gamma^{a'b'}\bigr)^{\beta\gamma}
	\Bigr]
\nn\\
	-sq\Bigl[
	\bigl(\gamma^{ab}\bigr)^{\alpha\delta}
	-\bigl(  &&\kern-25pt    \gamma^{a'b'} \bigr)^{\alpha\delta}
	\Bigr]
	-2\Bigl[q^2+r^2+s^2- 
	s(q+r)\Bigr] 
	\delta^{a'b'}\delta^{\alpha\delta}\delta^{\beta\gamma}
	 \biggr\}	\ .
\ee
The $H_{\rm BF}$ sector also makes the following contribution to the 
off-diagonal $(2,3)$ block:
\be
\Braket{J| b_q^{\alpha} b_r^{\beta} a_s^{A} (H_{\rm BF}) 
	a_r^{B\dag} b_q^{\gamma\dag} b_s^{\delta\dag}|J}
	=  
	-\frac{\lambda'}{2J}\delta^{\alpha\gamma}rs
	\biggl\{
	\bigl(\delta^{ab}-\delta^{a'b'}
	\bigr)\delta^{\beta\delta}
	-\bigl(\gamma^{ab}\bigr)^{\beta\delta}
	+\bigl(\gamma^{a'b'}\bigr)^{\beta\delta}
	\biggr\}\ .
\ee
The contributions of $H_{\rm BF}$ to the remaining off-diagonal blocks $(2,3)$, 
$(2,4)$, etc.~are obtained by appropriate index permutations.


The sector of spacetime fermions decomposes in a similar fashion.
The fermion analogue of table~\ref{boseblock3} for the bosonic sector
appears in table~\ref{fermiblock3}.
\begin{table}[ht!]
\begin{eqnarray}
\begin{array}{|c|cccc|}
\hline
 {H}_{\rm int} & 
	b_s^{\zeta\dag}  b^{\epsilon\dagger}_r  b^{\delta\dagger}_{q}\ket{J} &
	a_s^{C\dag}  a^{D\dagger}_r  b^{\delta\dagger}_{q}\ket{J} &
        a_r^{C\dag}  a^{D\dagger}_q  b^{\delta\dagger}_{s}\ket{J} &
	a_r^{C\dag}  a^{D\dagger}_s  b^{\delta\dagger}_{q}\ket{J}
\\   \hline
\bra{J} b^{\alpha}_q  b^{\beta}_r  b^{\gamma}_{s}      
				& H_{\rm FF}    & H_{\rm BF} & H_{\rm BF} & H_{\rm BF} \\
\bra{J} b^{\alpha}_q  a^{A}_r  a^{B}_{s}& 
		H_{\rm BF} & H_{\rm BB}+H_{\rm BF} & H_{\rm BF} & H_{\rm BF} \\
\bra{J} b^{\alpha}_s  a^{A}_q  a^{B}_{r}& 
		H_{\rm BF} & H_{\rm BF} & H_{\rm BB}+H_{\rm BF} & H_{\rm BF} \\ 
\bra{J} b^{\alpha}_r  a^{A}_s  a^{B}_{q}& 
		H_{\rm BF} & H_{\rm BF} & H_{\rm BF} & H_{\rm BB}+ {H}_{\rm BF}\\ 
\hline
\end{array} \nonumber
\end{eqnarray}
\caption{Interaction Hamiltonian on spacetime fermion 
three-impurity states $(q\neq r\neq s)$}
\label{fermiblock3}
\end{table}
The $(1,1)$ fermion block is occupied by the pure-fermion sector of the Hamiltonian
taken between the purely fermionic three-impurity states 
$b^{\alpha\dag}_q  b^{\beta\dag}_r  b^{\gamma\dag}_{s}\Ket{J}$:
\be
\Braket{J| b_q^{\alpha} b_r^{\beta} b_s^{\gamma} (H_{\rm FF}) 
	b_s^{\zeta\dag} b_r^{\epsilon\dag} b_q^{\delta\dag}|J}
	& = & 
	-\frac{\lambda'}{J}\left[
	q^2+r^2+s^2-rs-q(r+s)\right]\delta^{\alpha\delta}
	\delta^{\beta\epsilon}\delta^{\gamma\zeta}
\nn\\	+ \frac{\lambda'}{24J}\delta^{\alpha\delta}rs
	\biggl\{
	&&\kern-25pt
	\bigl(\gamma^{ij}\bigr)^{\beta\gamma} 
	\bigl(\gamma^{ij}\bigr)^{\epsilon\zeta} 
	- \bigl(\gamma^{ij}\bigr)^{\beta\epsilon}
	\bigl(\gamma^{ij}\bigr)^{\gamma\zeta}
	+ \bigl(\gamma^{ij}\bigr)^{\beta\zeta}
	\bigl(\gamma^{ij}\bigr)^{\gamma\epsilon}
\nn\\	- &&\kern-25pt  \bigl(\gamma^{i'j'}\bigr)^{\beta\gamma}
	\bigl(\gamma^{i'j'}\bigr)^{\epsilon\zeta}
	+ \bigl(\gamma^{i'j'}\bigr)^{\beta\epsilon}
	\bigl(\gamma^{i'j'}\bigr)^{\gamma\zeta}
	- \bigl(\gamma^{i'j'}\bigr)^{\beta\zeta}
	\bigl(\gamma^{i'j'}\bigr)^{\gamma\epsilon}
\nn\\	+  
	\bigl( &&\kern-25pt
	r\rightleftharpoons q,\ 
	\alpha\rightleftharpoons \beta,\ 
	\delta\rightleftharpoons \epsilon
	 \bigr)
	+  
	\bigl( 
	s\rightleftharpoons q,\ 
	\alpha\rightleftharpoons \gamma,\ 
	\delta\rightleftharpoons \zeta
	 \bigr)
	\biggr\}\ .
\ee
The off-diagonal $(1,2)$, $(1,3)$ and $(1,4)$ blocks (and their transposes)
mix purely fermionic states with $a_s^{A\dag} a_r^{B\dag} b_q^{\alpha\dag}\Ket{J}$
states:
\be
\Braket{J| b_q^{\alpha} b_r^{\beta} b_s^{\gamma} (H_{\rm BF}) 
	a_s^{A\dag} a_r^{B\dag} b_q^{\delta\dag}|J}
	& = & 
	-\frac{\lambda'}{2J}\delta^{\alpha\delta}rs
	\biggl\{
	\left(\gamma^{ab'}\right)^{\beta\gamma}
	- \left(\gamma^{a'b}\right)^{\beta\gamma}
	\biggr\}\ .
\label{12fermiblock}
\ee
The lower-diagonal $(2,2)$, $(3,3)$ and $(4,4)$ blocks receive contributions from the
pure boson sector of the Hamiltonian:
\be
\Braket{J| b_q^{\alpha} a_r^{A} a_s^{B} (H_{\rm BB}) 
	a_s^{C\dag} a_r^{D\dag} b_q^{\beta\dag}|J}
	& = & 
	-\frac{\lambda'}{2J}\delta^{\alpha\beta}
	\biggl\{
	(r-s)^2\delta^{BC}\delta^{AD}
	-(r^2+s^2)\Bigl(
	\delta^{ad}\delta^{bc}
	-\delta^{a'd'}\delta^{b'c'}
	\Bigr)
\nn\\	-2rs\Bigl(
	\delta^{ac}\delta^{bd} 
	-\delta^{ab}\delta^{cd}  &&\kern-25pt
	-\delta^{a'c'}\delta^{b'd'}
	+\delta^{a'b'}\delta^{c'd'}
	\Bigr)
	+(r^2-s^2)\Bigl(
	\delta^{ad}\delta^{b'c'}
	- \delta^{a'd'}\delta^{bc}
	\Bigr)
	\biggr\}\ .
\ee
In the same diagonal blocks of table~\ref{fermiblock3}, the $H_{\rm BF}$
sector contributes
\be
\Braket{J| b_q^{\alpha} a_r^{A} a_s^{B} (H_{\rm BF}) 
	a_s^{C\dag} a_r^{D\dag} b_q^{\beta\dag}|J}
	& = & 
	\frac{\lambda'}{8J}
	\biggl\{
	\delta^{\alpha\beta}\Bigl[
	+\bigl(
	8q(r+s)-5(r^2+s^2)-6q^2
	\bigr)\delta^{AD}\delta^{BC}
\nn\\
	+ (3q^2+s^2)\delta^{AD}\delta^{bc} &&\kern-25pt
	+ (3q^2+r^2)\delta^{BC}\delta^{ad} 
	+ (r^2-5q^2)\delta^{BC}\delta^{a'd'}
	+ (s^2-5q^2)\delta^{AD}\delta^{b'c'}
	\Bigr]
\nn\\	-4\delta^{BC}qr\Bigl[	&&\kern-25pt
	\bigl(\gamma^{ad}\bigr)^{\alpha\beta}
	- \bigl(\gamma^{a'd'}\bigr)^{\alpha\beta}
	\Bigr]
	-4\delta^{AD}qs\Bigl[
	\bigl(\gamma^{bc}\bigr)^{\alpha\beta}
	- \bigl(\gamma^{b'c'}\bigr)^{\alpha\beta}
	\Bigr]
	\biggr\}\ .
\ee
Finally, the off-diagonal blocks $(2,3)$, $(2,4)$ and $(3,4)$ (plus their
transpose entries) are given by the $H_{\rm BF}$ matrix element
\be
\Braket{J| b_q^{\alpha} a_r^{A} a_s^{B} (H_{\rm BF}) 
	a_r^{C\dag} a_q^{D\dag} b_s^{\beta\dag}|J}
	& = & 
	-\frac{\lambda'}{32 J}\delta^{AC}\biggl\{
	\delta^{\alpha\beta}
	\Bigl[
	(q-s)^2\delta^{BD}
	-(q^2+14qs+ s^2)\delta^{bd}
\nn\\	
	-( &&\kern-28pt    q^2-  18qs+  s^2)\delta^{b'd'}
	\Bigr]  
	+16 qs \Bigl[
	\bigl(\gamma^{bd}\bigr)^{\alpha\beta}
	-\bigl(\gamma^{b'd'}\bigr)^{\alpha\beta}
	\Bigr]
	\biggr\}\ .
\ee

A significant departure from the two-impurity case is that all these matrix
elements have, along with their spacetime index structures, non-trivial
dependence on the mode indices. The eigenvalues could potentially have very
complicated mode-index dependence but, as we shall see, they do not.
This amounts to a rigid consistency check on the whole procedure that 
was not present in the two-impurity case.

\subsection{Matrix diagonalization: inequivalent mode indices $(q\neq r\neq s)$}
We now turn to the task of diagonalizing the one-loop approximation
to the perturbing Hamiltonian. To simplify the task, we exploit certain 
block diagonalizations that hold to leading order in $\lambda'$ 
(but not to higher orders). While we eventually want to study the spectrum 
to all orders in $\lambda'$, diagonalizing the Hamiltonian at one loop
will reveal the underlying supermultiplet structure. As an example of the
simplifications we have in mind, we infer from (\ref{12block}) that the 
matrix elements of $H_{\rm BF}$ between pure boson states 
$a_q^{A\dag} a_r^{B\dag} a_s^{C\dag}\ket{J}$ and bifermionic spacetime bosons
vanish to leading order in $\lambda'$ if all three $SO(8)$ bosonic vector 
indices lie within the same $SO(4)$, descended either from 
$AdS_5$ or $S^5$. Restricting to such states brings 
the bosonic sector of the Hamiltonian into the block-diagonal form in 
table~\ref{boseblock1}.
\begin{table}[ht!]
\begin{eqnarray}
\begin{array}{|c|cc|}
\hline
 {H}_{\rm int} & a^{a\dagger} a^{b\dagger} a^{c\dagger} \ket{J}
		+ a^{a'\dagger} a^{b'\dagger} a^{c'\dagger} \ket{J}  &
	a^{A\dagger}  b^{\alpha\dagger}  b^{\beta\dagger}  \ket{J} 
\\   \hline
\bra{J} a^{a} a^{b} a^{c} +
\bra{J} a^{a'} a^{b'} a^{c'} & { H}_{\rm BB} & 0  \\
\bra{J} a^{A}  b^{\alpha}  b^{\beta} & 0  & H_{\rm FF}+H_{\rm BF}  \\
\hline
\end{array} \nonumber
\end{eqnarray}
\caption{Block-diagonal $SO(4)$ projection on bosonic three-impurity string states}
\label{boseblock1}
\end{table}
This leaves two 64-dimensional subspaces of purely bosonic states on which
the perturbation is block diagonal, as recorded in table~\ref{boseblock2}.  
\begin{table}[ht!]
\begin{eqnarray}
\begin{array}{|c|cc|}
\hline
 {H}_{\rm int} & a^{a\dagger} a^{b\dagger} a^{c\dagger} \ket{J} &
	a^{a'\dagger} a^{b'\dagger} a^{c'\dagger}\ket{J} 
\\   \hline
\bra{J} a^{a} a^{b} a^{c} & ({ H}_{\rm BB})_{64\times 64} & 0  \\
\bra{J} a^{a'} a^{b'} a^{c'} & 0  & (H_{\rm BB})_{64\times 64}  \\
\hline
\end{array} \nonumber
\end{eqnarray}
\caption{SO(4) projection on purely bosonic states}
\label{boseblock2}
\end{table}

Since the interaction Hamiltonian has manifest $SO(4)\times SO(4)$ 
symmetry, it is useful to project matrix elements onto irreps of that group
before diagonalizing. In some cases the irrep is unique, and
projection directly identifies the corresponding eigenvalue. In the cases
where an irrep has multiple occurrences, there emerges an unavoidable matrix 
diagonalization that is typically of low dimension. In what follows,
we will collect the results of carrying out this program on the one-loop
interaction Hamiltonian.  A very important feature of the results which
appear is that all the eigenvalues turn out to have a common simple 
dependence on mode indices. More precisely, 
the expansion of the eigenvalues for inequivalent mode indices $(q,r,s)$ 
out to first non-leading order in $\lambda'$ and $1/J$ can be written as
\be
\label{lambdaexp}
E_J(q,r,s) = 3+\frac{\lambda'(q^2+r^2+s^2)}{2}
	\left(1 + \frac{\Lambda}{J} + O(J^{-2})\right)\ ,
\ee
where $\Lambda$ is a pure number that characterizes the lifting of the 
degeneracy in the various sectors. The notation $\Lambda_{\rm BB},\ \Lambda_{\rm BF}$ and 
$\Lambda_{\rm FF}$ will be used to denote energy corrections arising entirely from 
the indicated sectors of the perturbing Hamiltonian. This simple quadratic 
dependence of the eigenvalues on the mode indices does not automatically 
follow from the structure of the matrix elements themselves, but is
important for the successful match to gauge theory eigenvalues. 
In what follows, we will catalog some of the different $\Lambda$ values that 
occur, along with their $SO(4)\times SO(4)$ irreps (and multiplicities). 
When we have the complete list, we will discuss 
how they are organized into supermultiplets.

In the $SO(4)$ projection in table~\ref{boseblock2}, we will find a set 
of 64 eigenvalues for both the $SO(4)_{AdS}$ and $SO(4)_{S^5}$ subsectors.
We record this eigenvalue spectrum in table~\ref{SO4BOSE}, using an 
$SU(2)^2\times SU(2)^2$ notation. For comparison, it is displayed alongside
the projection of the 2-impurity spectrum onto the same subspace (as 
found in \cite{Callan:2003xr,STRpaper}).
\begin{table}[ht!]
\begin{equation}
\begin{array}{|c|c|}
\hline
SO(4)_{AdS}\times SO(4)_{S^5} & \Lambda_{\rm BB} \\
\hline 
{\bf (1,1;2,2)}	&	-8	\\
{\bf [1,1;(2+4),2]}+{\bf [1,1;2,(2+4)]}	&	-6	\\
{\bf [1,1;(2+4),(2+4)]}	&	-4 \\
\hline
{\bf [(2+4),(2+4);1,1]}	&	-2	\\
{\bf [(2+4),2;1,1]}+{\bf [2,(2+4);1,1]}	&	0	\\
{\bf (2,2;1,1)}	&	2 \\
\hline
\end{array} \nonumber \qquad\qquad
\begin{array}{|c|c|}
\hline
SO(4)_{AdS}\times SO(4)_{S^5} & \Lambda_{\rm BB} \\
\hline 
{\bf (1,1;1,1)}	&	-6	\\
{\bf (1,1;3,1)}+{\bf (1,1;1,3)}	&	-4	\\
{\bf (1,1;3,3)}	&	-2 \\
\hline
{\bf (3,3;1,1)}	&	-2	\\
{\bf (3,1;1,1)}+{\bf (1,3;1,1)}	&	0	\\
{\bf (1,1;1,1)}	&	2 \\
\hline
\end{array} \nonumber
\end{equation}
\caption{Three-impurity energy spectrum in the pure-boson $SO(4)$ projection
(left panel) and two-impurity energy spectrum in the same projection (right panel)}
\label{SO4BOSE}
\end{table}
In the three-impurity case, the ${\bf (1,1;2,2)}$ level in the
$SO(4)_{S^5}$ subsector clearly descends from the two-impurity
singlet ${\bf (1,1;1,1)}$ in the same $SO(4)$ subgroup.  
In the same manner, the three-impurity ${\bf [1,1;(2+4),2]}+{\bf [1,1;2,(2+4)]}$
level descends from the $SO(4)_{S^5}$ antisymmetric two-impurity state
${\bf (1,1;3,1)}+{\bf (1,1;1,3)}$, and the three-impurity ${\bf [1,1;(2+4),(2+4)]}$
level is tied to the two-impurity symmetric-traceless ${\bf (1,1;3,3)}$ irrep.
In the $SO(4)_{S^5}$ subsector, each of these levels receives a shift to the 
energy of $-2$.  The total multiplicity of each of these levels is also increased 
by a factor of four when the additional ${\bf (2,2)}$ is tensored into the 
two-impurity state space.  The $SO(4)_{AdS}$ subsector follows a similar pattern:
the ${\bf (2,2;1,1)}$, ${\bf [(2+4),2;1,1]}+{\bf [2,(2+4);1,1]}$ and 
${\bf [(2+4),(2+4);1,1]}$
levels appear as three-impurity descendants of the two-impurity irrep spectrum 
${\bf (1,1;1,1)}+{\bf (3,1;1,1)}+{\bf (1,3;1,1)}+{\bf (3,3;1,1)}$.
In this subsector, however, the three-impurity energies are identical 
to those in the two-impurity theory.  

The bosonic $SO(4)$ projection has a precise fermionic analogue.  
Similar to the bosons, the $SO(9,1)$ spinors $b_q^\dag$ decompose 
as ${\bf (2,1;2,1)}+{\bf (1,2;1,2)}$ under the action of $\Pi$ parity:
\be
\label{fermipi}
\Pi \hat b_q^\dag = \hat b_q^\dag \qquad 
\Pi \tilde b_q^\dag = - \tilde b_q^\dag\ .
\ee
The notation $\hat b_q^\dag$ labels ${\bf (1,2;1,2)}$ spinors with
positive eigenvalue under $\Pi$, and $\tilde b_q^\dag$ 
indicate ${\bf (2,1;2,1)}$ spinors which are negative under $\Pi$.
Analogous to the $SO(4)$ projection on the $SO(8)$ bosonic operators 
$a_q^{A\dag } \to a_q^{a\dag} + a_q^{a'\dag}$, 
projecting out the positive or negative eigenvalues of $\Pi$ on the 
eight-component spinor $b_q^{\alpha\dag}$ 
leaves a subspace of four-component spinors spanned by 
$\hat b_q^\dag$ and $\tilde b_q^\dag$.

We can perform a projection on the subsector in table~\ref{boseblock3}
similar to that appearing in table~\ref{boseblock2}.
In this case, instead of three bosonic impurities mixing with
a single bosonic (plus a bi-fermionic) excitation,
we are now interested in projecting out particular interactions
between a purely fermionic state and a state with one fermionic 
and two bosonic excitations. 
Using $\pm$ to denote the particular representation of the 
fermionic excitations, the off-diagonal elements given 
by (\ref{12fermiblock}) vanish for $+++ \to \pm$ and $---\to \pm$ interactions.
In other words, the pure fermion states in the $(1,1)$ block 
of table~\ref{fermiblock3} will not 
mix with states containing two bosonic excitations if 
all three fermionic oscillators lie in the same $\Pi$ projection.  
This projection appears schematically in table~\ref{fermiblock1}.
\begin{table}[ht!]
\begin{eqnarray}
\begin{array}{|c|cc|}
\hline
 {H}_{\rm int} & \hat b^{\alpha\dagger}\hat b^{\beta\dagger}\hat b^{\gamma\dagger} \ket{J} 
	+ \tilde b^{\alpha\dagger}\tilde b^{\beta\dagger}\tilde b^{\gamma\dagger} 
	\ket{J}& 
	a^{A\dagger} a^{B\dagger} b^{\alpha\dagger} \ket{J}  
		\\   \hline
\bra{J}\hat b^{\alpha}\hat b^{\beta}\hat b^{\gamma}+
\bra{J}\tilde b^{\alpha}\tilde b^{\beta}\tilde b^{\gamma}	  
					& { H}_{\rm FF} & 0  \\
\bra{J} a^{A} a^{B} b^{\alpha}	 & 0  & H_{\rm BB}+H_{\rm BF}  \\
\hline
\end{array} \nonumber
\end{eqnarray}
\caption{Block-diagonal projection on fermionic three-impurity string states}
\label{fermiblock1}
\end{table}

The $(1,1)$ pure fermion block in table~\ref{fermiblock1} 
breaks into two 64-dimensional subsectors under this
projection. By tensoring an additional ${\bf (1,2;1,2)}$ or
${\bf (2,1;2,1)}$ impurity into the two-impurity
state space, we expect to see a multiplicity structure
in this projection given by
\be
{\bf (1,2)}\times {\bf(1,2;1,2)} & = & 
	{\bf (1,2;1,2)} + {\bf [1,2;1,(2+4)]} 
\nn\\
&&	+ {\bf [1,(2+4);1,2]}
	+ {\bf [1,(2+4);1,(2+4)]}\ ,
\nn\\
{\bf (2,1)}\times {\bf(2,1;2,1)} & = & 
	{\bf (2,1;2,1)} + {\bf [2,1;(2+4),1]} 
\nn\\
&&	+ {\bf [(2+4),1;2,1]}
	+ {\bf [(2+4),1;(2+4),1]}\ , 
\ee
for a total of 128 states. The projections onto the two 64-dimensional 
$\Pi_+$ and $\Pi_-$ subspaces yield identical eigenvalues and multiplicities. 
The results for both subspaces are presented in table~\ref{Pifermi}:
\begin{table}[ht!]
\begin{equation}
\begin{array}{|c|c|}
\hline
SO(4)_{AdS}\times SO(4)_{S^5} & \Lambda_{\rm FF} \\
\hline 
{\bf (2,1;2,1)}+{\bf (1,2;1,2)}	&	-3	\\
{\bf [2,1;(2+4),1]}+{\bf [1,2;1,(2+4)]}	&	-1	\\
{\bf [(2+4),1;2,1]}+{\bf [1,(2+4);1,2]}	&	-5 \\
{\bf [(2+4),1;(2+4),1]}+{\bf [1,(2+4);1,(2+4)]}	&	-3 \\
\hline
\end{array} \nonumber
\qquad\qquad
\begin{array}{|c|c|}
\hline
SO(4)_{AdS}\times SO(4)_{S^5} & \Lambda_{\rm FF} \\
\hline 
{\bf (1,1;1,1)}+{\bf (1,1;1,1)}	&	-2	\\
{\bf (1,1;3,1)}+{\bf (1,1;1,3)}	&	0	\\
{\bf (3,1;1,1)}+{\bf (1,3;1,1)}	&	-4 \\
{\bf (3,1;3,1)}+{\bf (1,3;1,3)}	&	-2 \\
\hline
\end{array} \nonumber
\end{equation}
\caption{Spectrum of three-impurity states (left panel) and two-impurity 
states (right panel) created by $\Pi_\pm$-projected fermionic creation operators}
\label{Pifermi}
\end{table}
The two-impurity bi-fermion states in table~\ref{Pifermi} 
are spacetime bosons while the tri-fermion states
are spacetime fermions. For comparison purposes, we have displayed
both spectra. Note that the $O(1/J)$ energy corrections of the 
two types of state are simply displaced by $-1$ relative to each other.

This exhausts the subspaces that can be diagonalized by simple 
irrep projections. The remaining eigenvalues must be obtained by
explicit diagonalization of finite dimensional submatrices obtained
by projection onto representations with multiple occurrence. The upshot
of these more complicated eigenvalue calculations is that the first-order 
$\lambda'$ eigenvalues take on all integer values from $\Lambda=-8$ to
$\Lambda= +2$, alternating between spacetime bosons and fermions as 
$\Lambda$ is successively incremented by one unit. 

\subsection{Assembling eigenvalues into supermultiplets}

Finally, we need to understand how the perturbed three-impurity
spectrum breaks up into extended supersymmetry multiplets. This is
relatively easy to infer from the multiplicities of the perturbed
eigenvalues (and the multiplicities are a side result of the calculation 
of the eigenvalues themselves). In the last subsection, we described a 
procedure for diagonalizing the one-loop
perturbing Hamiltonian on the $4,096$-dimensional space of 
three-impurity string states with mode indices $p\ne q\ne r$. 
The complete results for the eigenvalues $\Lambda$ and their 
multiplicities are stated in table~\ref{multlist} (we use the notation of 
(\ref{lambdaexp}), while the $B$ and $F$ subscripts are used to indicate 
bosonic and fermionic levels in the supermultiplet).
\begin{table}[ht!]
\begin{eqnarray}
\begin{array}{|c|ccccccccccc|}
\hline
\Lambda	& -8	& -7	& -6    & -5	& -4	& -3	& -2 & -1 & 0 
		& 1 & 2 \\	
\hline
{\rm Multiplicity}	 & 4_B	& 40_F	& 180_B	& 480_F	& 840_B	& 1008_F
		& 840_B	& 480_F	& 180_B	& 40_F	& 4_B	 \\
\hline
\end{array} \nonumber
\end{eqnarray}
\caption{Complete three-impurity energy spectrum (with multiplicities)}
\label{multlist}
\end{table}

The eigenvalues $\Lambda$ must meet certain conditions if the requirements of 
$PSU(2,2|4)$ symmetry are to be met. The eigenvalues in question are lightcone
energies and thus dual to the gauge theory quantity $\Delta=D-J$, the difference
between scaling dimension and ${R}$-charge. Since conformal invariance is part
of the full symmetry group, states are organized into conformal multiplets built 
on conformal primaries. A supermultiplet will contain several conformal primaries 
having the same value of $\Delta$ and transforming into each other under the supercharges. 
All 16 supercharges increment the dimension of an operator by $1/2$, but only 8 
of them (call them ${\cal Q}_\alpha$) also increment the ${R}$-charge by $1/2$, so as to
leave $\Delta$ unchanged. These 8 supercharges act as `raising operators' on the 
conformal primaries of a supermultiplet: starting from a super-primary of lowest
${R}$-charge, the other conformal primaries are created by acting on it in all 
possible ways with the eight ${\cal Q}_\alpha$. Primaries obtained by acting with $L$ factors 
of ${\cal Q}_\alpha$ on the super-primary are said to be at level $L$ in the supermultiplet 
(since the  ${\cal Q}_\alpha$ anticommute, the range is $L=0$ to $L=8$). The multiplicities 
of states at the various levels are also determined: for each $L=0$ primary operator, there 
will be $C^8_L$ such operators at level $L$ (where $C^n_m$ is the binomial coefficient). 
If the $L=0$ primary has multiplicity $s$, summing over all $L$ gives $2^8 s=256 s$ 
conformal primaries in all. 

These facts severely restrict the quantity $\Lambda$ in the general expression 
(\ref{lambdaexp}) above. Although the states in the degenerate multiplet all
have the same $J$, they actually belong to different levels $L$ in more than one 
supermultiplet.  A state of given $L$ is a member of a supermultiplet 
built on a `highest-weight' or super-primary state with ${ R}=J-L/2$. 
Since all the primaries in a supermultiplet have the same $\Delta$, the joint 
dependence of eigenvalues on $\lambda,J,L$ must be of the form 
$\Delta(\lambda,J-L/2)$. The only way the expansion of (\ref{lambdaexp}) 
can be consistent with this is if $\Lambda=L+c$, where $c$ is a pure numerical 
constant (recall that $\lambda'=\lambda/J^2$). Successive members of
a supermultiplet must therefore have eigenvalues separated by exactly $1$
and the difference between `top' ($L=8$) and `bottom' ($L=0$) eigenvalues for
$\Lambda$ must be exactly $8$. 

The $\Lambda$ eigenvalues in table~\ref{multlist} are integer-spaced, which
is consistent with supersymmetry requirements. However, because the range 
between top and bottom eigenvalues is $10$, rather than $8$, the $4,096$-dimensional 
space must be built on more than one type of extended supermultiplet, with more
than one choice of $c$ in the general formula $\Lambda=L+c$. 
This is to be contrasted with the two-impurity case, where the degenerate space 
was exactly $256$-dimensional and was spanned by a single superconformal
primary whose lowest member was a singlet under both Lorentz transformations 
and the residual $SO(4)$ ${R}$-symmetry. We can readily infer what 
superconformal primaries are needed to span the degenerate three-impurity
state space by applying a little numerology to table~\ref{multlist}.
The lowest eigenvalue is $\Lambda=-8$: it has multiplicity $4$ and,
according to table~\ref{SO4BOSE}, its $SO(4)\times SO(4)$ decomposition
is ${\bf (1,1;2,2)}$ (spacetime scalar, ${R}$-charge $SO(4)$ four-vector).
According to the general arguments about how the full extended supermultiplet
is built by acting on a `bottom' state with the eight raising operators,
it is the base of a supermultiplet of $4\times 256$ states extending
up to $\Lambda=0$. By the same token, there is a highest eigenvalue
$\Lambda=+2$: it has multiplicity $4$ and, according to table~\ref{SO4BOSE}, 
its $SO(4)\times SO(4)$ decomposition is ${\bf (2,2;1,1)}$ (spacetime vector, 
${R}$-charge singlet). Using lowering operators instead of
raising operators, we see that one derives from it a supermultiplet of
$4\times 256$ operators with eigenvalues extending from $\Lambda=-6$
to $\Lambda=+2$. The multiplicities of the $\Lambda$ eigenvalues occurring
in these two supermultiplets are of course given by binomial coefficients,
as described above.  By comparing with the \emph{total} multiplicities of 
each allowed $\Lambda$ (as listed in table~\ref{multlist}) we readily see
that what remains are $8\times 256$ states with eigenvalues running from 
$\Lambda=-7$ to $\Lambda=+1$ with the correct binomial coefficient pattern 
of multiplicities. The top and bottom states here are spacetime fermions 
and must lie in a spinor representation of the Lorentz group. It is not 
hard to see that they lie in the eight-dimensional $SO(4)\times SO(4)$ irrep 
${\bf (2,1;1,2)}+{\bf (1,2;2,1)}$. This exhausts all the states and we conclude that 
the three-impurity state space is spanned by three distinct extended 
superconformal multiplets. The detailed spectrum is given in table~\ref{3mult1} 
(where the last line records the total multiplicity at each level as given 
in table~\ref{multlist} and the first line records the two-impurity spectrum
for reference). Note the peculiar feature that certain energies are shared
by all three multiplets: this is an accidental degeneracy that does not
survive at higher loop order.
\begin{table}[ht!]
\begin{eqnarray}
\begin{array}{|c|ccccccccccc|c|}
\hline
\Lambda	& -8	& -7	& -6    & -5	& -4	& -3	& -2	
		& -1	& 0	& 1	& 2	& 	\\ \hline\hline
\Delta_0=2	& 	& 	& 1_B	& 8_F	& 28_B	& 56_F	& 70_B	
		& 56_F	& 28_B	& 8_F	& 1_B	& {\rm scalar}	\\ \hline\hline 
\Delta_0=3	 & 4 & 32	& 112	& 224	& 280	& 224	& 112	
		& 32	& 4	& 	& 	& SO(4)_{S^5} ~{\rm vector} \\ \hline
	& 	& 	& 4 	& 32	& 112	& 224	& 280	
		& 224	& 112	& 32	& 4	& SO(4)_{AdS_5} ~{\rm vector} \\ \hline
	& 	& 8	& 64	& 224	& 448	& 560	& 448 	
		& 224	& 64	& 8	& 	& {\rm spinor}	\\ \hline
\hline
{\rm Total}	 & 4_B	& 40_F	& 180_B	& 480_F	& 840_B	& 1008_F
		& 840_B	& 480_F	& 180_B	& 40_F	& 4_B	& 4,096 \\
\hline
\end{array} \nonumber
\end{eqnarray}
\caption{Submultiplet breakup of the three-impurity spectrum}
\label{3mult1}
\end{table}

A complete analysis of the agreement with gauge theory anomalous 
dimensions will have to be deferred until a later 
section: the dimensions of three-impurity gauge theory operators 
are much harder to calculate than those of the two-impurity operators
and there are few results in the literature, even at one loop. However, 
it is worth making a few preliminary remarks at this point. Since there 
are three superconformal multiplets, we have only three independent 
anomalous dimensions to compute. Minahan and Zarembo \cite{Minahan:2002ve} 
found that the problem simplifies dramatically if we study the one-loop 
anomalous dimension of the special subset of single-trace operators of the form
$\tr{(\phi^I Z^J)}$ (and all possible permutations of the fields inside the
trace), where the $R$-charge is carried by an $SO(4)\times SO(4)$ 
singlet scalar field $Z$ and the impurities are insertions of a scalar field 
$\phi$ lying in the ${\bf (1,1;2,2)}$ (vector) irrep of the residual 
$SO(4)\times SO(4)$ symmetry.  More formally, these operators
are in the $SO(4)\times SO(4)$ irrep obtained by completely symmetrizing $I$
vectors in the ${\bf (1,1;2,2)}$ irrep. The crucial point is that such operators 
form a `closed sector', mixing \emph{only} among themselves under the anomalous 
dimension operator.  More importantly, the action of the one-loop anomalous 
dimension operator on this closed sector can be recast as the action of an integrable spin 
chain Hamiltonian of a type solvable by Bethe ansatz techniques. Although the Bethe ansatz 
is generally not analytically soluble, Minahan and Zarembo used it to 
obtain a virial expansion for the anomalous dimension in which the number $I$ of 
impurities is held fixed, while the $R$-charge $J$ is taken to be large
(see eqn.~(5.29) in \cite{Minahan:2002ve}).  In terms of the number of
spin chain lattice sites ${\alg L}$, their result appears as
\be
\gamma_{\so(6)} = \frac{\lambda}{2{\alg L}^3}\sum_n M_n k_n^2 
	\left( {\alg L} + M_n + 1 \right) + O({\alg L}^{-4})\ .
\ee
The integer $k_n$ represents pseudoparticle momenta on the spin chain, 
and is dual to the string theory worldsheet mode indices;
the quantity $M_n$ labels the number of trace impurities with identical $k_n$.
With $I$ impurities, the spin chain length is given in terms of the $R$-charge
by ${\alg L} = J+I$, which leads to
\be
\gamma_{\so(6)} = \frac{\lambda}{2J^3} \sum_n M_n k_n^2
	\left( J - 2I +M_n + 1 \right) + O(J^{-4})\ .
\label{MZBA}
\ee
This virial expansion
is similar in character to (\ref{lambdaexp}) and, for $I=3$ (the three-impurity 
case), it matches that equation precisely with $\Lambda=-4$.

On the string theory side, three completely symmetrized 
${\bf (1,1;2,2)}$ vectors form a tensor in the ${\bf (1,1;4,4)}$ irrep; such 
an irrep can be constructed from three $SO(4)_{S^5}$ vector (bosonic) creation operators.
Table~\ref{SO4BOSE} shows that the corresponding string perturbation theory 
eigenvalue is (at one-loop order) $\Lambda=-4$ as well. We infer from 
table~\ref{3mult1} that this eigenvalue lies at level $L=4$ of the $SO(4)_{S^5}$ 
vector superconformal multiplet (and this argument takes care of the gauge 
theory/string theory comparison for all other operators in that multiplet). 

The sector described above is often 
called an $\so(6)$\footnote{This notation
is used to distinguish the protected gauge theory symmetry groups from those in the string theory.}   
sector on the gauge theory side, with reference to the subalgebra of the 
full superconformal algebra under which it is invariant.
In an $\su(2)$ subspace of the $\so(6)$, this sector becomes closed to all loop order.  
For future reference, we note that Beisert \cite{Beisert:2003jj} has identified two other 
`closed sectors' of operators in the gauge theory.  In addition to the bosonic $\su(2)$ sector,
a bosonic $\Sl(2)$ sector and an $\su(2|3)$ sector (of which the closed $\su(2)$ sector is a subgroup) 
are also exactly closed.  
It should be noted that integrable $\Sl(2)$ spin chains were discovered some time ago in 
phenomenologically-motivated studies of the scaling behavior of high-energy scattering 
amplitudes in physical, non-supersymmetric QCD \cite{Belitsky:1999bf} 
(see also \cite{Kotikov:2000pm,Kotikov:2001sc,Kotikov:2002ab,Kotikov:2003fb}).
The $\su(2|3)$ spin chain was studied more recently in \cite{Beisert:2003ys}:  
this closed sector breaks into the $\su(2)$ bosonic sector and a fermionic subsector 
which, to avoid confusion, we simply denote as a subgroup of $\su(2|3)$.

In the string theory, 
the subsectors analogous to the gauge theory $\Sl(2)$ and $\su(2|3)$ 
are constructed out of
completely symmetrized $SO(4)_{AdS}$ bosons and completely symmetrized fermions 
of the same $\Pi$ eigenvalue, respectively \cite{STRpaper}.  
They correspond to the central $L=4$ levels
of the remaining two supermultiplets in table~\ref{3mult1}, and a calculation
of their eigenvalues would complete the analysis of the match between 
three-impurity operators and string states at one-loop order. 
Unfortunately, explicit general
results for three-impurity operator dimensions, analogous to those obtained 
by Minahan and Zarembo for the $\so(6)$ sector \cite{Minahan:2002ve}, have not
been obtained for the other closed sectors.  The Bethe ansatz for the
general one-loop integrable spin chain presented in \cite{Beisert:2003yb} 
could easily be exploited for this purpose.\footnote{We are indebted to N.~Beisert 
for this observation.} 
However, since we eventually want to go beyond one-loop, where Bethe 
ansatz technology is less well-developed, we have found it more useful to develop 
numerical methods for evaluating spin chain eigenvalues (we refer the reader to 
\cite{spinchain} for a check of our results against Bethe-ansatz techniques,
including the higher-loop corrections of \cite{Beisert:2004hm}).  
This subject will be developed in a later section.


\subsection{Two equivalent mode indices $(q=r=n,~s=-2n)$}

When two mode indices are allowed to be equal, the analysis becomes
slightly more complicated. Since we are diagonalizing a
Hamiltonian that is quartic in oscillators in a basis of 
three-impurity string states, one oscillator in the ``in'' state must 
always be directly contracted with one oscillator in the ``out'' state
and, with two equal mode indices, there are many more nonvanishing 
contributions to each matrix element. While the matrix 
elements are more complicated, the state space is only half as large when
two mode indices are allowed to be equal (only half as many mode-index 
permutations on the basis states generate linearly independent states).  
As a result, the fermionic and bosonic sectors of the Hamiltonian are 
are each 1,024-dimensional. By the same token, the multiplet structure
of the energy eigenstates will be significantly different from the
unequal mode index case studied in the previous subsection.  

To study this case, we make the mode index choice
\be
q=r=n \qquad s=-2n\ .
\ee 
The structure of matrix elements of the string Hamiltonian between spacetime bosons 
is given in table~\ref{boseblock3nn}. 
\begin{table}[ht!]
\begin{eqnarray}
\begin{array}{|c|ccc|}
\hline
 {H}_{\rm int} & 
	a_{-2n}^{D\dag}  a^{E\dagger}_n  a^{F\dagger}_{n}\ket{J} &
	a_{-2n}^{D\dag}  b^{\gamma\dagger}_n  b^{\delta\dagger}_{n}\ket{J} &
        a_n^{D\dag}  b^{\gamma\dagger}_n  b^{\delta\dagger}_{-2n}\ket{J} 
\\   \hline
\bra{J} a^{A}_n  a^{B}_n  a^{C}_{-2n}      & H_{\rm BB}    
		& H_{\rm BF} & H_{\rm BF}  \\
\bra{J} a^{A}_n  b^{\alpha}_n  b^{\beta}_{-2n}& H_{\rm BF} 
		& H_{\rm FF}+H_{\rm BF} & H_{\rm BF}  \\
\bra{J} a^{A}_{-2n}  b^{\alpha}_n  b^{\beta}_{n}& H_{\rm BF} 
		& H_{\rm BF} & H_{\rm FF}+H_{\rm BF}  \\ 
\hline
\end{array} \nonumber
\end{eqnarray}
\caption{Bosonic three-impurity string perturbation matrix with $(q= r=n,\  s=-2n)$}
\label{boseblock3nn}
\end{table}
This table seems to describe a $3\times 3$ block matrix with $512\times 512$ blocks 
in each subsector, giving a 1,536-dimensional state space.  However, the vector and spinor 
indices are required to run over values that generate linearly independent basis 
states. This eliminates one third of the possible index assignments, implying that
the matrix is in fact $1,024\times 1,024$. 

To evaluate the entries in table~\ref{boseblock3nn}, we express the Hamiltonians 
(\ref{HBBfinal}-\ref{HBFfinal}) in terms of mode creation and annihilation operators, 
expand the result in powers of $\lambda'$ and compute the indicated matrix 
elements between three-impurity Fock space states. We collect below all the relevant 
results of this exercise for this equal-mode-index case.

The purely bosonic subsector in the $(1,1)$ block is given by
\be
\Bra{J}&&\kern-20pt
	a_n^{A} a_n^{B} a_{-2n}^{C} (H_{\rm BB})a_{-2n}^{D\dag} a_n^{E\dag} 
	a_n^{F\dag}\Ket{J}
	 = 
\frac{n^2\,\lambda}{2J} \biggl\{
	5\,{\delta }^{BF}\,\delta^{cd}\delta^{ae} + 
      5\,{\delta }^{AF}\,\delta^{cd}\delta^{be} 
	- 4\,{\delta }^{BF}\,\delta^{ad}\delta^{ce} 
\nn\\
&&	+     4\,{\delta }^{BF}\,\delta^{ac}\delta^{de} 
	+ 4\,{\delta }^{AF}\,\delta^{bc}\delta^{de} +
      5\,{\delta }^{BE}\,\delta^{cd}\delta^{af} 
	- 4\,{\delta }^{BE}\,\delta^{ad}\delta^{cf} + 
      4\,{\delta }^{BE}\,\delta^{ac}\delta^{df} 
	+ 4\,{\delta }^{AE}\,\delta^{bc}\delta^{df}
\nn\\
&&	-       4\,{\delta }^{bd}
	\Bigl( {\delta}^{AF}\,{\delta }^{ce} 
	+ {\delta }^{AE}\,{\delta }^{cf}
         \Bigr)  		
	+ 3\,{\delta }^{BF}\,{\delta }^{ae}\,{\delta }^{c'd'} 
	+ 3\,{\delta }^{AF}\,{{{\delta }}}^{be}\,{{{\delta }}}^{c'd'} + 
      3\,{\delta }^{BE}\,{{{\delta }}}^{af}\,{{{\delta }}}^{c'd'} 
\nn\\
&&	-      3\,{\delta }^{BF}\,{{{\delta }}}^{cd}\,{{{\delta }}}^{a'e'}  
    	-3\,{\delta }^{AF}\,{{{\delta }}}^{cd}\,{{{\delta }}}^{b'e'}   
	- 5\,{\delta }^{BF}\,{{{\delta }}}^{c'd'}\delta^{a'e'} - 
      5\,{\delta }^{AF}\,{{{\delta }}}^{c'd'}\delta^{b'e'} 
	+ 4\,{\delta }^{BF}\,{{{\delta }}}^{a'd'}\delta^{c'e'} 
\nn\\
&&	+       4\,{\delta }^{AF}\,{{{\delta }}}^{b'd'}\delta^{c'e'} 
	- 4\,{\delta }^{BF}\,{{{\delta }}}^{a'c'}     \delta^{d'e'} - 
      4\,{\delta }^{AF}\,{{{\delta }}}^{b'c'}\delta^{d'e'} 
	- 3\,{\delta }^{BE}\,{{{\delta }}}^{cd}\,{{{\delta }}}^{a'f'} - 
     3\,{\delta }^{AE}\,{{{\delta }}}^{cd}\,{{{\delta }}}^{b'f'} 
\nn\\
&&	- 5\,{\delta }^{BE}\,{{{\delta }}}^{c'd'}\delta^{a'f'} - 
     5\,{\delta }^{AE}\,{{{\delta }}}^{c'd'}          \delta^{b'f'}     
	+ 4\,{\delta }^{BE}\,{{{\delta }}}^{a'd'}\delta^{c'f'} + 
      4\,{\delta }^{AE}\,{{{\delta }}}^{b'd'}\delta^{c'f'} 
	- 4\,{\delta }^{BE}\,{{{\delta }}}^{a'c'}\delta^{d'f'} 
\nn\\
&&	-       4\,{\delta }^{AE}\,{{{\delta }}}^{b'c'}\delta^{d'f'}  
    	+  {\delta }^{AE}\,{{{\delta }}}^{bf}\,  
	\Bigl( 5\,{{{\delta }}}^{cd} 
	+ 3\,{{{\delta }}}^{c'd'} \Bigr)  - 
      2\,{\delta }^{CD}\,\Bigl[ 9\,
	\Bigl( {\delta }^{BE}\delta^{AF} 
	+ {\delta }^{AE}\delta^{BF} \Bigr)  
	-          {{{\delta }}}^{be}\delta^{af} 
\nn\\
&&	- {{{\delta }}}^{ae}\delta^{bf} 
	+ {{{\delta }}}^{ab}\delta^{ef} +       
         {{{\delta }}}^{b'e'}\delta^{a'f'} + {{{\delta }}}^{a'e'}\delta^{b'f'} 
	- {{{\delta }}}^{a'b'}\delta^{e'f'} 
	\Bigr] 
	\biggr\}\ .
\label{bosonsnn}
\ee
This matrix element exhibits the same antisymmetry between the $SO(4)_{AdS}$ 
and $SO(4)_{S^5}$ indices that is exhibited in eqn.~(\ref{bosons}).  
The off-diagonal $H_{\rm BF}$ mixing sector is essentially equivalent to its
counterpart in eqn.~(\ref{12block}):
\be
\Braket{J| a_n^{A} a_n^{B} a_{-2n}^{C} (H_{\rm BF}) 
	a_{-2n}^{D\dag} b_n^{\alpha\dag} b_n^{\beta\dag}|J}
	& = & 
	\frac{n^2\lambda'}{2J}\delta^{CD}
	\biggl\{
	\left(\gamma^{ab'}\right)^{\alpha\beta} - 
	\left(\gamma^{a'b}\right)^{\alpha\beta}
	\biggr\}\ .
\label{12blocknn}
\ee
The diagonal contributions from the pure fermion sector $H_{\rm FF}$
in the $(2,2)$ and $(3,3)$ blocks of table~\ref{boseblock3nn} appear as
\be
\Braket{J| b_n^{\alpha} b_n^{\beta} a_{-2n}^{A} (H_{\rm FF}) 
	a_{-2n}^{B\dag} b_n^{\gamma\dag} b_n^{\delta\dag}|J}
	& = & 
	\frac{n^2\lambda'}{24 J}\delta^{AB} \biggl\{
	\bigl(\gamma^{ij}\bigr)^{\alpha\gamma} 
	\bigl(\gamma^{ij}\bigr)^{\beta\delta}
	- \bigl(\gamma^{ij}\bigr)^{\alpha\beta}
	\bigl(\gamma^{ij}\bigr)^{\gamma\delta}
\nn\\
	- \bigl(\gamma^{ij}\bigr)^{\alpha\delta}
	\bigl(\gamma^{ij}\bigr)^{\beta\gamma}  
	-\bigl(\gamma^{i'j'}\bigr)^{\alpha\gamma} 
	\bigl(   &&\kern-30pt  \gamma^{i'j'} \bigr)^{\beta\delta} 
	+ \bigl(\gamma^{i'j'}\bigr)^{\alpha\beta}
	\bigl(\gamma^{i'j'}\bigr)^{\gamma\delta}
	+ \bigl(\gamma^{i'j'}\bigr)^{\alpha\delta}
	\bigl(\gamma^{i'j'}\bigr)^{\beta\gamma}
	\biggr\}\ .
\label{HFF22nn}
\ee
The $H_{\rm BF}$ sector exhibits the following contribution
to the lower diagonal blocks $(2,2)$ and $(3,3)$:
\be
\Braket{J| b_n^{\alpha} b_n^{\beta} a_{-2n}^{A} (H_{\rm BF}) 
	a_{-2n}^{B\dag} b_n^{\gamma\dag} b_n^{\delta\dag}|J}
	& = & 
	\frac{n^2\lambda'}{J}\biggl\{
	-10\,\delta^{a'b'} \left(
	\delta^{\alpha\delta}\delta^{\beta\gamma}
	-\delta^{\alpha\gamma}\delta^{\beta\delta}
	\right)
\nn\\
	-8\,\delta^{ab} \left(
	\delta^{\alpha\delta}\delta^{\beta\gamma}  
	-\delta^{\alpha\gamma}\delta^{\beta\delta}
	\right)				&&\kern-25pt
	-\delta^{\alpha\gamma}
	\Bigl[
	(\gamma^{ab})^{\beta\delta} - (\gamma^{a'b'})^{\beta\delta}
	\Bigr]
	+\delta^{\alpha\delta}
	\Bigl[
	(\gamma^{ab})^{\beta\gamma} - (\gamma^{a'b'})^{\beta\gamma}
	\Bigr]
\nn\\
	+\delta^{\beta\gamma}
	\Bigl[					&&\kern-25pt
	(\gamma^{ab})^{\alpha\delta} - (\gamma^{a'b'})^{\alpha\delta}
	\Bigr]
	-\delta^{\beta\delta}
	\Bigl[
	(\gamma^{ab})^{\alpha\gamma} - (\gamma^{a'b'})^{\alpha\gamma}
	\Bigr]
	\biggr\}\ .
\label{HBF2211}
\ee
Finally, the off-diagonal version of (\ref{HBF2211}) appears in the $(2,3)$ block
(along with its transpose in the $(3,2)$ block): 
\be
\Braket{J| b_n^{\alpha} b_n^{\beta} a_{-2n}^{A} (H_{\rm BF}) 
	a_n^{B\dag} b_n^{\gamma\dag} b_{-2n}^{\delta\dag}|J}
	& = &
	-\frac{n^2\lambda'}{J}\biggl\{
	\delta^{a'b'} \left(
	\delta^{\alpha\delta}\delta^{\beta\gamma}
	-\delta^{\alpha\gamma}\delta^{\beta\delta}
	\right)
	-\delta^{ab} \left(
	\delta^{\alpha\delta}\delta^{\beta\gamma}  
	-\delta^{\alpha\gamma}\delta^{\beta\delta}
	\right)				
\nn\\
	+\delta^{\alpha\gamma}
	\Bigl[  &&\kern-25pt
	(\gamma^{ab})^{\beta\delta} - (\gamma^{a'b'})^{\beta\delta}
	\Bigr]		
	-\delta^{\beta\gamma}
	\Bigl[				
	(\gamma^{ab})^{\alpha\delta} - (\gamma^{a'b'})^{\alpha\delta}
	\Bigr]
	\biggr\}\ .
\ee

The fermionic sector perturbation matrix is displayed schematically 
in table~\ref{fermiblock3nn}. Like table~\ref{boseblock3nn}, it is 
$1,024\times 1,024$ once redundant index assignments are eliminated.
\begin{table}[ht!]
\begin{eqnarray}
\begin{array}{|c|ccc|}
\hline
 {H}_{\rm int} & 
	b_{-2n}^{\zeta\dag}  b^{\epsilon\dagger}_n  b^{\delta\dagger}_{n}\ket{J} &
	a_{-2n}^{C\dag}  a^{D\dagger}_n  b^{\delta\dagger}_{n}\ket{J} &
        a_n^{C\dag}  a^{D\dagger}_n  b^{\delta\dagger}_{-2n}\ket{J} 
\\   \hline
\bra{J} b^{\alpha}_n  b^{\beta}_n  b^{\gamma}_{-2n}      
						& H_{\rm FF}    
		& H_{\rm BF} & H_{\rm BF}  \\
\bra{J} b^{\alpha}_n  a^{A}_n  a^{B}_{-2n}& H_{\rm BF} 
		& H_{\rm BB}+H_{\rm BF} & H_{\rm BF}  \\
\bra{J} b^{\alpha}_{-2n}  a^{A}_n  a^{B}_{n}& H_{\rm BF} & H_{\rm BF} 
		& H_{\rm BB}+H_{\rm BF}  \\ 
\hline
\end{array} \nonumber
\end{eqnarray}
\caption{Fermionic three-impurity string perturbation matrix $(q=r=n,\ s=-2n)$}
\label{fermiblock3nn}
\end{table}

The purely fermionic subsector in the $(1,1)$ block of table~\ref{fermiblock3nn}
takes the form
\be
\Braket{J| b_n^{\alpha} b_n^{\beta} b_{-2n}^{\gamma} (H_{\rm FF}) 
	b_{-2n}^{\zeta\dag} b_n^{\epsilon\dag} b_n^{\delta\dag}|J}
	& = & 
	\frac{9n^2\lambda'}{J}\delta^{\gamma\zeta}\left(
	\delta^{\alpha\epsilon}\delta^{\beta\delta} 
	- \delta^{\alpha\delta}\delta^{\beta\epsilon}
	\right)
\nn\\	+\frac{n^2\lambda'}{24J}\biggl\{
	\delta^{\gamma\zeta}\Bigl[    
	(\gamma^{ij})^{\alpha\beta}(\gamma^{ij})^{\delta\epsilon} &&\kern-25pt
	- (\gamma^{ij})^{\alpha\delta}(\gamma^{ij})^{\beta\epsilon}
	+ (\gamma^{ij})^{\alpha\epsilon}(\gamma^{ij})^{\beta\delta}
	-(\gamma^{i'j'})^{\alpha\beta}(\gamma^{i'j'})^{\delta\epsilon}  
\nn\\	+ (\gamma^{i'j'})^{\alpha\delta}(\gamma^{i'j'})^{\beta\epsilon} &&\kern-25pt
	- (\gamma^{i'j'})^{\alpha\epsilon}(\gamma^{i'j'})^{\beta\delta}  
	\Bigr]
	-2\delta^{\alpha\delta}\Bigl[
	(\gamma^{ij})^{\beta\gamma}(\gamma^{ij})^{\epsilon\zeta}
	- (\gamma^{ij})^{\beta\epsilon}(\gamma^{ij})^{\gamma\zeta}
\nn\\	+ (\gamma^{ij})^{\beta\zeta}(\gamma^{ij})^{\gamma\epsilon}  &&\kern-25pt
	-(\gamma^{i'j'})^{\beta\gamma}(\gamma^{i'j'})^{\epsilon\zeta}
	+ (\gamma^{i'j'})^{\beta\epsilon}(\gamma^{i'j'})^{\gamma\zeta}
	- (\gamma^{i'j'})^{\beta\zeta}(\gamma^{i'j'})^{\gamma\epsilon}
	\Bigr]
\nn\\	+2\delta^{\alpha\epsilon}\Bigl[
	(\gamma^{ij})^{\beta\gamma}(\gamma^{ij})^{\delta\zeta}  &&\kern-25pt
	- (\gamma^{ij})^{\beta\delta}(\gamma^{ij})^{\gamma\zeta}
	+ (\gamma^{ij})^{\beta\zeta}(\gamma^{ij})^{\gamma\delta}
	-(\gamma^{i'j'})^{\beta\gamma}(\gamma^{i'j'})^{\delta\zeta}
\nn\\	+ (\gamma^{i'j'})^{\beta\delta}(\gamma^{i'j'})^{\gamma\zeta}  &&\kern-25pt
	- (\gamma^{i'j'})^{\beta\zeta}(\gamma^{i'j'})^{\gamma\delta}
	\Bigr]
	+2\delta^{\beta\delta}\Bigl[
	(\gamma^{ij})^{\alpha\gamma}(\gamma^{ij})^{\epsilon\zeta}
	- (\gamma^{ij})^{\alpha\epsilon}(\gamma^{ij})^{\gamma\zeta}
\nn\\	+ (\gamma^{ij})^{\alpha\zeta}(\gamma^{ij})^{\gamma\epsilon}  &&\kern-25pt
	-(\gamma^{i'j'})^{\alpha\gamma}(\gamma^{i'j'})^{\epsilon\zeta}
	+ (\gamma^{i'j'})^{\alpha\epsilon}(\gamma^{i'j'})^{\gamma\zeta}
	- (\gamma^{i'j'})^{\alpha\zeta}(\gamma^{i'j'})^{\gamma\epsilon}
	\Bigr]
\nn\\	-2\delta^{\beta\epsilon}\Bigl[
	(\gamma^{ij})^{\alpha\gamma}(\gamma^{ij})^{\delta\zeta}  &&\kern-25pt
	- (\gamma^{ij})^{\alpha\delta}(\gamma^{ij})^{\gamma\zeta}
	+ (\gamma^{ij})^{\alpha\zeta}(\gamma^{ij})^{\gamma\delta}
	-(\gamma^{i'j'})^{\alpha\gamma}(\gamma^{i'j'})^{\delta\zeta} 
\nn\\	+ (\gamma^{i'j'})^{\alpha\delta}(\gamma^{i'j'})^{\gamma\zeta}  &&\kern-25pt
	- (\gamma^{i'j'})^{\alpha\zeta}(\gamma^{i'j'})^{\gamma\delta}
	\Bigr]
	\biggr\}\ .
\label{pureferminn}
\ee
The off-diagonal blocks $(1,2)$ and $(1,3)$ receive contributions
from the $H_{\rm BF}$ sector:
\be
\Braket{J| b_n^{\alpha} b_n^{\beta} b_{-2n}^{\gamma} (H_{\rm BF}) 
	a_{-2n}^{A\dag} a_n^{B\dag} b_n^{\delta\dag}|J}
	& = & 
	\frac{n^2\lambda'}{J}\biggl\{
	\delta^{\alpha\delta}\Bigl[
	\left(\gamma^{ab'}\right)^{\beta\gamma}
	- \left(\gamma^{a'b}\right)^{\beta\gamma}
	\Bigr]
\nn\\
&&	-\delta^{\delta\beta}\Bigl[
	\left(\gamma^{ab'}\right)^{\alpha\gamma}
	- \left(\gamma^{a'b}\right)^{\alpha\gamma}
	\Bigr] \biggr\}\ .
\label{12fermiblocknn}
\ee
The bosonic sector $H_{\rm BB}$ contributes to the $(2,2)$ and $(3,3)$ blocks:
\be
\Braket{J| b_q^{\alpha} a_r^{A} a_s^{B} (H_{\rm BB}) 
	a_s^{C\dag} a_r^{D\dag} b_q^{\beta\dag}|J}
	& = & 
	-\frac{n^2\lambda'}{2J}\delta^{\alpha\beta}\biggl\{
	9\,\delta^{AD}\delta^{BC}
	+4\,\delta^{ac}\delta^{bd}
	-4\,\delta^{ab}\delta^{cd}
\nn\\	-\delta^{ad}\left(
		5\,\delta^{bc}+3\,\delta^{b'c'}\right)  &&\kern-25pt
	-4\,\delta^{a'c'}\delta^{b'd'}
	+4\,\delta^{a'b'}\delta^{c'd'}
	+\delta^{a'd'}\left(
		5\,\delta^{b'c'}+3\,\delta^{bc}\right)
	\biggr\}\ .
\ee
In the same lower-diagonal blocks, $H_{\rm BF}$ exhibits the contribution
\be
\Braket{J| b_n^{\alpha} a_n^{A} a_{-2n}^{B} (H_{\rm BF}) 
	a_{-2n}^{C\dag} a_n^{D\dag} b_n^{\beta\dag}|J}
	& = & 
	-\frac{n^2\lambda'}{8J}\biggl\{
	39\,\delta^{\alpha\beta}\delta^{AD}\delta^{BC}
	+\delta^{\alpha\beta}\delta^{AD}
	\left(\delta^{b'c'}-7\,\delta^{bc}\right)
\nn\\	-4\,\delta^{\alpha\beta}\delta^{BC}
	\left(\delta^{ad}-\delta^{a'd'}\right) &&\kern-25pt
	+4\,\delta^{BC}\Bigl[
	(\gamma^{ad})^{\alpha\beta}
	-(\gamma^{a'd'})^{\alpha\beta}
	\Bigr]
	-8\,\delta^{AD}\Bigl[  
	(\gamma^{bc})^{\alpha\beta}
	-(\gamma^{b'c'})^{\alpha\beta}
	\Bigr]
	\biggr\}\ .
\nn\\
&&
\ee
Finally, $H_{\rm BF}$ yields matrix elements in the off-diagonal block
$(2,3)$:
\be
\Braket{J| b_n^{\alpha} a_n^{A} a_{-2n}^{B} (H_{\rm BF}) 
	a_n^{C\dag} a_n^{D\dag} b_{-2n}^{\beta\dag}|J}
	& = & 
	-\frac{n^2\lambda'}{32J}\biggl\{
	9\,\delta^{\alpha\beta}\delta^{AC}\delta^{BD}
	+9\,\delta^{\alpha\beta}\delta^{AD}\delta^{BC}
\nn\\
	+\delta^{\alpha\beta}       \delta^{AC} &&\kern-25pt   \left(  
	23\delta^{bd}-41\delta^{b'd'}\right) 
	+\delta^{\alpha\beta}\delta^{AD}\left(
	23\delta^{bc}-41\delta^{b'c'}\right)
\nn\\
	-32\delta^{AD}   \Bigl[     
	(\gamma^{bc}   )^{\alpha\beta}   &&\kern-25pt  
	- (\gamma^{b'c'})^{\alpha\beta}
	\Bigr]
	-32\delta^{AC}\Bigl[
	(\gamma^{bd})^{\alpha\beta} 
	- (\gamma^{b'd'})^{\alpha\beta}
	\Bigr]
	\biggr\}\ .
\ee

We can perform a full symbolic diagonalization of the $1,024\times 1,024$ bosonic
and fermionic perturbation matrices to obtain the one-loop in $\lambda'$, $O(1/J)$ 
energy corrections. They can all be expressed in terms of dimensionless
eigenvalues $\Lambda$ according to the standard formula (\ref{lambdaexp}) modified
by setting $q=r=n,\ s=-2n$:
\be
E_J(n) = 3+3n^2\lambda'\left(1+\frac{\Lambda}{J}+O(J^{-2})\right)\ .
\label{Lexpnn}
\ee
The resulting spectrum is displayed in table~\ref{nnspec}.
\begin{table}[ht!]
\begin{eqnarray}
\begin{array}{|c|ccccccccc|}
\hline
\Lambda_1~({S^5~{\rm vector}})
	& -23/3& -20/3	& -17/3	& -14/3	& -11/3	& -8/3 & -5/3	& -2/3	& 1/3	 \\
\hline
{\rm Multiplicity}	& 4_B	& 32_F	& 112_B    & 224_F & 280_B & 224_F	
								& 112_B	& 32_F	& 4_B 	\\
\hline\hline
\Lambda_2~({AdS_5~{\rm vector}})	
	& -19/3	& -16/3	& -13/3	& -10/3	& -7/3	& -4/3 & -1/3	& 2/3	& 5/3	 \\
\hline
{\rm Multiplicity}	& 4_B	& 32_F	& 112_B    & 224_F & 280_B & 224_F	
								& 112_B	& 32_F	& 4_B 	\\
\hline
\end{array} \nonumber
\end{eqnarray}
\caption{Spectrum of three-impurity string Hamiltonian with $(q=r=n,\ s=-2n)$}
\label{nnspec}
\end{table}
The levels clearly organize themselves into two superconformal multiplets built on 
vector primary states. Note that the spinor multiplet is absent and that the degeneracy 
between multiplets that was seen in the inequivalent mode index case has been lifted. 
The spinor multiplet is absent for the following reason: it contains a representation 
at level $L=4$ arising from fermion creation operators completely symmetrized on 
$SO(4)\times SO(4)$ spinor indices; such a construct must vanish unless all the creation 
operator mode indices are different. 

If we keep track of the $SO(4)\times SO(4)$
irrep structure, we find that the symmetric-traceless bosonic $SO(4)_{S^5}$ states 
arising from the closed $\su(2)$ subsector fall into the $-11/3~[280_B]$ level. This 
is the counterpart of the $-4~[280_B]$ level in table~\ref{3mult1}. To compare
with Minahan and Zarembo's Bethe ansatz calculation of the corresponding gauge theory 
operator dimension, we must evaluate eqn.~(\ref{MZBA}) with the
appropriate choice of parameters.  In particular, $M_n=2$ 
when two mode indices are allowed to coincide and, comparing with eqn.~(\ref{Lexpnn}),
we find perfect agreement with the string theory prediction $\Lambda=-11/3$. 
States at level $L=4$ in the second multiplet in table~\ref{nnspec} correspond
to operators in the $\Sl(2)$ closed sector of the gauge theory and the eigenvalue
$\Lambda=-7/3~[280_B]$ amounts to a prediction for the one-loop anomalous
dimension of that class of gauge theory operators. As mentioned at the end of the
previous subsection, we will need to develop a numerical treatment of the
$\Sl(2)$ spin chain Hamiltonian in order to assess the agreement between 
string theory and gauge theory in this sector. 

\section{Three impurity string spectrum: all orders in $\lambda'$}

In the previous section, we have studied the eigenvalue spectrum of the 
string theory perturbation Hamiltonian expanded to leading order in $1/J$ 
and to one-loop order in $\lambda'$. The expansion in $\lambda'$ was for
convenience only since our expressions for matrix elements are exact in
this parameter.  We should, in principle, be able to obtain results that are 
exact in $\lambda'$ (but still of leading order in $1/J$). This is a
worthwhile enterprise since recent progress on the gauge theory side
has made it possible to evaluate selected operator anomalous dimensions
to two- and three-loop order. The simple one-loop 
calculations of the previous sections have given us an overview
of how the perturbed string theory eigenvalues are organized into 
superconformal multiplets. This provides a very useful orientation for 
the more complex all-orders calculation, to which we now turn.

\subsection{Inequivalent mode indices: $(q\neq r\neq s)$}

Our first step is to collect the exact matrix elements of the perturbing
Hamiltonian between three-impurity states of unequal mode indices. The 
block structure of the perturbation matrix in the spacetime boson sector
is given in table~\ref{boseblock3} and the exact form of the $(1,1)$ block is
\be
\Bra{J}&&\kern-20pt a_q^{A} a_r^{B} a_{s}^{C} (H_{\rm BB})a_{s}^{D\dag} a_r^{E\dag} 
	a_q^{F\dag}\Ket{J}
	 = 
	-\frac{1}{2\omega_q\omega_r\omega_s}\biggl\{
	\delta^{BE}\omega_r
	\Bigl[
	\delta^{CD}\delta^{AF}(s^2+q^2(1+2s^2\lambda'))
\nn\\
&&	- (q^2+s^2)\delta^{cd}\delta^{af}	
	-2qs (\delta^{ad}\delta^{cf}  		
	-\delta^{ac}\delta^{df})
	+(q^2-s^2)\delta^{af}\delta^{c'd'}
	-(q^2-s^2)\delta^{a'f'}\delta^{cd}
\nn\\
&&	+ (q^2+s^2)\delta^{c'd'}\delta^{a'f'}	
	+2qs (\delta^{a'd'}\delta^{c'f'}  	
	-\delta^{a'c'}\delta^{d'f'})
	\Bigr]
	+ \Bigl(
	C\rightleftharpoons B,\ 	
	D\rightleftharpoons E,\		
	s\rightleftharpoons r
	\Bigr)
\nn\\
&&	+ \Bigl(
	A\rightleftharpoons B,\ 
	F\rightleftharpoons E,\
	q\rightleftharpoons r
	\Bigr)
	\biggr\}\ ,
\label{fullbose11}
\ee
where we define $\omega_q \equiv \sqrt{q^2+1/\lambda'}$ to simplify this and 
other similar expressions.

The off-diagonal $H_{\rm BF}$ contributions to the $(1,2)$, $(1,3)$ and $(1,4)$ blocks
are yet more complicated.  To simplify the expressions, we define
\be
F_1 \equiv \sqrt{(\omega_q+q)(\omega_r-r)} & \qquad & 
		F_2 \equiv \sqrt{(\omega_q-q)(\omega_r+r)} \nn\\
F_3 \equiv \sqrt{(\omega_q-q)(\omega_r-r)} & \qquad & 
		F_4 \equiv \sqrt{(\omega_q+q)(\omega_r+r)}\ .
\ee
Using these functions, the matrix elements in these off-diagonal
subsectors are given by:
\be
\Bra{J}&&\kern-20pt a_q^{A} a_r^{B} a_{s}^{C} (H_{\rm BF}) 
	a_{s}^{D\dag} b_r^{\alpha\dag} b_q^{\beta\dag}\Ket{J}
	 = 
	\frac{\delta^{CD}}{32 \omega_q\omega_r J}\biggl\{
	\frac{8}{\sqrt{\lambda'}}
	(F_1-F_2)\delta^{AB}\delta^{\gamma\delta}  
	-2(q-r)(F_3+F_4)\delta^{AB}\delta^{\gamma\delta} 
\nn\\
&&	+4(q-r)(F_3+F_4)(\gamma^{ab})^{\gamma\delta}		
	-2(q+r)(F_3-F_4)(\gamma^{ab'})^{\gamma\delta}
\nn\\
&&	+(2qF_3 -2qF_4 +2rF_3 - 2rF_4 )(\gamma^{a'b})^{\gamma\delta}
	-(4qF_3+4qF_4-4rF_3-4rF_4) (\gamma^{a'b'})^{\gamma\delta} 
\nn\\
&&	+\frac{8}{\sqrt{\lambda'}}(F_2-F_1)\delta^{a'b'}  \delta^{\gamma\delta}  
	+4(q- r)   
	(F_3+F_4)\delta^{\gamma\delta}\delta^{a'b'}
	-2(q-r)(F_3+F_4)\delta^{\gamma\delta}(\delta^{ab}-\delta^{a'b'})
\nn\\
&&	-4\lambda'\omega_q\omega_r(q-r)(F_3+   F_4)\delta^{AB}\delta^{\gamma\delta}
	+4\sqrt{\lambda'}(qr-\omega_q\omega_r)
	\Bigl[
	(F_1+F_2)\Bigl(
	(\gamma^{ab'})^{\gamma\delta} - (\gamma^{a'b})^{\gamma\delta}
	\Bigr)
\nn\\
&&	-(F_1-F_2)\delta^{\gamma\delta}(\delta^{ab}-  \delta^{a'b'})
	\Bigr]   
	+2(\omega_q+\omega_r)(F_3+F_4)
	\Bigl[
	(\gamma^{ab'})^{\gamma\delta} - (\gamma^{a'b})^{\gamma\delta}
	\Bigr]
\nn\\
&&	+4\sqrt{\lambda'}(r\omega_q-q\omega_r)(F_1+F_2)
	\Bigl[  
	(\gamma^{ab})^{\gamma\delta}- (\gamma^{a'b'})^{\gamma\delta}
	\Bigr]
\nn\\
&&	-4\lambda'(q-r)(F_3-F_4)(r\omega_q+q\omega_r)\delta^{AB}\delta^{\gamma\delta}
	-\lambda'\delta^{AB}
\nn\\
&&	+2\lambda'(\omega_q\omega_r-qr)(q-r)(F_3+F_4)\delta^{AB}\delta^{\gamma\delta}
	+4\sqrt{\lambda'}(\omega_q\omega_r+qr)(F_1-F_2)\delta^{AB}\delta^{\gamma\delta}
\nn\\
&&	-2\lambda'  (q-r)(\omega_q\omega_r+qr)  (F_3+F_4)\delta^{AB}\delta^{\gamma\delta}
	\biggr\}\ .
\ee
The $H_{\rm FF}$ contribution to the lower-diagonal blocks $(2,2)$, $(3,3)$ and
$(4,4)$ is
\be
\Bra{J}&&\kern-20pt b_q^{\alpha} b_r^{\beta} a_{s}^{A} (\rm H_{\rm FF}) 
	a_{s}^{B\dag} b_r^{\gamma\dag} b_q^{\delta\dag}\Ket{J}
	 =
\nn\\
&&	\frac{\delta^{AB}}{48\omega_r\omega_s J}\sqrt{\lambda'}
	\biggl\{
	2 rs \sqrt{1/\lambda'}
	\biggl[
	\Bigl(
	(\gamma^{ij})^{\alpha\gamma}(\gamma^{ij})^{\beta\delta}
	-(\gamma^{i'j'})^{\alpha\gamma}(\gamma^{i'j'})^{\beta\delta}
	\Bigr)
\nn\\
&&	-\Bigl(
	(\gamma^{ij})^{\alpha\delta}(\gamma^{ij})^{\beta\gamma}
	-(\gamma^{i'j'})^{\alpha\delta}(\gamma^{i'j'})^{\beta\gamma}
	\Bigr)
	-\Bigl(
	(\gamma^{ij})^{\alpha\beta}(\gamma^{ij})^{\gamma\delta}
	-(\gamma^{i'j'})^{\alpha\beta}(\gamma^{i'j'})^{\gamma\delta}
	\Bigr)
	\biggr]
\nn\\
&&	-12
	\Bigl[
	2\delta^{\alpha\delta}\delta^{\beta\gamma}	
	\Bigl(
	s^2\sqrt{1/\lambda'}-2rs\sqrt{\lambda'}\omega_r\omega_s
	+r^2(2s^2\sqrt{\lambda'}+\sqrt{1/\lambda'})
	\Bigr) \Bigr]
	\biggr\}\ .
\ee
The bose-fermi Hamiltonian $H_{\rm BF}$ contributes the following matrix
elements to the same lower-diagonal blocks:
\be
\Bra{J}&&\kern-20pt b_q^{\alpha} b_r^{\beta} a_{s}^{A} (H_{\rm BF}) 
	a_{s}^{B\dag} b_r^{\gamma\dag} b_q^{\delta\dag}\Ket{J}
	 = 
	-\frac{1}{2\omega_q\omega_r\omega_s}
	\biggl\{
	s\sqrt{\lambda'}\delta^{ab}\delta^{\alpha\delta}\delta^{\beta\gamma}
	\Bigl[
	s\omega_r
	( 2q^2\sqrt{\lambda'}+\sqrt{1/\lambda'} )
\nn\\
&&	+ s\omega_q
	( 2r^2\sqrt{\lambda'}+\sqrt{1/\lambda'} )
	-2\omega_q\omega_r\omega_s(q+r)\sqrt{\lambda'}
	\Bigr]
	+\delta^{a'b'}\delta^{\alpha\delta}\delta^{\beta\gamma}
	\Bigl[
	2\omega_r q^2 (1+s^2\lambda') 
\nn\\
&&	+ s^2\omega_r
	+2\omega_q r^2 (1+s^2\lambda') 
	+ s^2\omega_q
	-2s(q+r)\lambda'\omega_q\omega_r\omega_s
	\Bigr]
\nn\\
&&	+sr\omega_q\delta^{\alpha\delta}\Bigl[
	(\gamma^{ab})^{\beta\gamma} - (\gamma^{a'b'})^{\beta\gamma}
	\Bigr]
	+sq\omega_r\delta^{\beta\gamma}\Bigl[
	(\gamma^{ab})^{\alpha\delta} - (\gamma^{a'b'})^{\alpha\delta}
	\Bigr]
	\biggr\}\ .
\nn\\
&&
\ee

To simplify off-diagonal elements in the $(2,3)$, $(2,4)$ and $(3,4)$
blocks, we define
\be
G_1 \equiv \sqrt{(\omega_r+r)(\omega_s-s)} & \qquad & 
		G_2 \equiv \sqrt{(\omega_r-r)(\omega_s+s)} \nn\\
G_3 \equiv \sqrt{(\omega_r-r)(\omega_s-s)} & \qquad & 
		G_4 \equiv \sqrt{(\omega_r+r)(\omega_s+s)}\ .
\ee
The matrix elements in these subsectors are then given by
\be
\Bra{J}&&\kern-20pt b_q^{\alpha} b_r^{\beta} a_{s}^{A} (H_{\rm BF}) 
	a_r^{B\dag} b_q^{\gamma\dag} b_{s}^{\delta\dag}\Ket{J}
	 = 
\nn\\
&&	-\frac{1}{16 (\lambda'\omega_r\omega_s)^{3/2}}\biggl\{
	\sqrt{\omega_r\omega_s}\lambda'
	\delta^{\alpha\gamma}
	\biggl[
	2\delta^{ab}\delta^{\beta\delta}\Bigl[
	(G_1+G_2)(2-2\lambda'\omega_r\omega_s)
\nn\\
&&	+(r+s)\sqrt{\lambda'}
	\Bigl(
	G_4-\lambda'G_4(r-\omega_r)(s-\omega_s)
	+G_3(-1+rs\lambda'+r\omega_s\lambda'+\omega_r(s+\omega_s)\lambda')
	\Bigr)
	\Bigr]
\nn\\
&&	+2\sqrt{\lambda'}
	\Bigl[
	(r+s)(G_3-G_4)+\sqrt{\lambda'}(G_1-G_2)(r\omega_s-s\omega_r)
	\Bigr]
	\Bigl[
	(\gamma^{ab})^{\beta\delta}
	-(\gamma^{a'b'})^{\beta\delta}
	\Bigr]
\nn\\
&&	+\sqrt{\lambda'}\Bigl[
	2rs\sqrt{\lambda'}G_1-2rs\sqrt{\lambda'}G_2+(r-s)(G_3+G_4)
	+(\omega_s-\omega_r)(G_3-G_4)
\nn\\
&&	+2\omega_r\omega_s\sqrt{\lambda'}(G_2-G_1)
	\Bigr]
	\Bigl[
	(\gamma^{ab'})^{\beta\delta}
	-(\gamma^{a'b})^{\beta\delta}
	\Bigr]
	+2\delta^{a'b'}\delta^{\beta\delta}
	\Bigl[
	-2rs\sqrt{\lambda'}(G_1-G_2)
\nn\\
&&	+(r+s)\sqrt{\lambda'}
	\Bigl(
	-G_4-\lambda'G_4(r-\omega_r)(s-\omega_s)
\nn\\
&&	+G_3(1+rs\lambda'+r\omega_s\lambda'+\omega_r(s+\omega_s)\lambda')
	\Bigr)
	\Bigr] \biggr]
	\biggr\}\ .
\ee

The entries in the spacetime fermion block matrix of table~\ref{fermiblock1} 
are far too complicated to write out explicitly: they
are best generated, viewed and manipulated with computer algebra
techniques. The explicit formulas, along with a collection of the
Mathematica programs written to generate and work with them, are 
available on the web \cite{webref}.

We were not able to symbolically diagonalize the complete perturbation
matrix built from the exact (in $\lambda'$) matrix elements listed
above: with the computing resources available to us, the routines for
diagonalizing the full 2,048-dimensional matrices would not terminate in 
any reasonable time. As noted in the previous section, however, gauge 
theory arguments suggest that there are three protected $SO(4)\times SO(4)$ 
irreps that do not mix with any other irreps. It is a straightforward
matter to project the perturbation matrix onto these unique protected 
irreps to obtain analytic expressions for the corresponding exact 
eigenvalues. In fact, the superconformal multiplet structure of the three-impurity 
problem is such that the energies/dimensions of all other irreps can be inferred from those 
of the three protected irreps. Hence, this method will give us exact
expressions for all the energy levels of the three-impurity problem.

Consider first the $\Sl(2)$ closed sector. The dual sector is generated on the 
string theory side by bosonic creation operators completely symmetrized (and traceless) 
on $SO(4)_{AdS}$ vector indices. 
The simplest way to make this projection on eqn.~(\ref{fullbose11}) is to 
compute diagonal elements between the symmetrized states
\be
a_q^{( a\dag}a_r^{b\dag}a_s^{c\dag )}\Ket{J}\ ,
\ee
with $a\neq b\neq c$ (and, of course, $a,b,c \in 1,\dots,4$).  
The charges of the fermionic oscillators 
under this subgroup are $\pm 1/2$, so the three-boson state of this type 
cannot mix with one boson and two fermions (or any other state). Hence,
the above projection of eqn.~(\ref{fullbose11}) yields the 
closed sector eigenvalue correction
\be
\delta E_{AdS} (q,r,s,J) & =  &
	\frac{1}{J\omega_q\omega_r\omega_s}\biggl\{
	qs(1-qs\lambda')\omega_r
	+qr(1-qr\lambda')\omega_s
	+rs(1-rs\lambda')\omega_q
\nn\\
&&\kern+60pt	+\left[
	qr+s(q+r)\right]\lambda'
	\omega_q\omega_r\omega_s
	\biggr\}
\nn\\
	&\approx & \frac{1}{J}\biggl\{
	-2(q^2+qr+r^2)\lambda'
	-\frac{15}{8}\bigl(q^2r^2(q+r)^2\bigr){\lambda'}^3
	+ \dots
	\biggr\}\ .
\label{exactSO41}
\ee
To facilitate eventual comparison with gauge theory results,
we have performed a small-$\lambda'$ expansion in the final line
with the substitution $s\to -(q+r)$ (since the mode indices
satisfy the constraint $s+q+r=0$). The leading correction 
$-2(q^2+qr+r^2)\lambda'$ reproduces the one-loop eigenvalue
$\Lambda_{\rm BB} = -2~[280_B]$ located at level $L=4$ in the $SO(4)_{AdS}$ 
multiplet in table~\ref{3mult1}. 


The closed $\su(2)$ sector is generated by bosonic creation operators
completely symmetrized on traceless $SO(4)_{S^5}$ indices. Projection onto this
irrep is most simply achieved by choosing all mode operators in 
eqn.~(\ref{fullbose11}) to carry symmetrized, traceless $SO(4)_{S^5}$ labels (they can
also be thought of as carrying charge $+1$ under some $SO(2)$ subgroup 
of $SO(4)_{S^5}$).  Direct projection yields the $SO(4)_{S^5}$ eigenvalue
\be
\delta E_{S^5}(q,r,s,J)  & = &
	-\frac{1}{J\omega_q\omega_r\omega_s}\biggl\{
	\left[qr+r^2+q^2(1+r^2\lambda')\right]\omega_s
	+ \left[qs+s^2+q^2(1+s^2\lambda')\right]\omega_r
\nn\\
&&	+ \left[rs+s^2+r^2(1+s^2\lambda')\right]\omega_q
	-\left[rs+q(r+s)\right]\lambda'\omega_q\omega_r\omega_s
	\biggr\}
\nn\\
&&\kern-30pt \approx 
	\frac{1}{J}\biggl\{
	-4(q^2+qr+r^2)\lambda'
	+(q^2+qr+r^2)^2{\lambda'}^2
\nn\\
&&\kern-20pt	-\frac{3}{4}\bigl(
	q^6+3q^5r+8q^4r^2+11q^3r^3
	+8q^2r^4+3qr^5+r^6\bigr)
	{\lambda'}^3 +\dots \biggr\}\ .
\label{exactSO42}
\ee
This is the all-loop formula corresponding 
to gauge theory operator dimensions in the closed $\su(2)$ subsector;
the leading-order term $-4(q^2+qr+r^2)\lambda'$ reproduces the 
one-loop eigenvalue $\Lambda_{\rm BB}=-4~[280_B]$ at level $L=4$
in the $SO(4)_{S^5}$ vector multiplet in table~\ref{3mult1}.  

The eigenvalue of the symmetrized pure-fermion irrep can be obtained 
by evaluating the exact matrix element $H_{\rm FF}$ acting on three 
symmetrized fermionic creation operators with $SO(4)\times SO(4)$ 
indices chosen to lie in the same $\Pi$ projection
(with inequivalent mode indices). The exact energy shift 
for this irrep turns out to be
\be
\delta E_{\rm Fermi}(q,r,s,J)  &=& 
	-\frac{1}{4\,J\omega_q\omega_r\omega_s}\biggl\{
	-4\bigl(rs+q(r+s)\bigr)\lambda'\omega_q\omega_r\omega_s
\nn\\
&&\kern-40pt	+\biggl[
	\omega_q
	\left(
	2 s^2
	+4r^2s^2\lambda'
	+2 r^2
	\right)
	+\bigl(s\to r,\ r\to q,\ q\to s\bigr)
	+\bigl( q\rightleftharpoons r \bigr)
	\biggr]
	\biggr\}
\nn\\
&&\kern-80pt \approx 
	\frac{1}{J}\biggl\{
	-3(q^2+qr+r^2)\lambda'
	+\frac{1}{2}(q^2+qr+r^2)^2{\lambda'}^2
\nn\\
&&\kern-70pt	-\frac{3}{16}\bigl(
	2q^6+6q^5r+21q^4r^2+32q^3r^3
	+21q^2r^4+6qr^5+2r^6\bigr)
	{\lambda'}^3 +\dots \biggr\}\ .
\label{exactfermi}
\ee
The leading-order $\lambda'$ correction $-3(q^2+qr+r^2)\lambda'$
reproduces the $\Lambda_{\rm FF}=-3~[580_F]$ eigenvalue at the $L=4$ level 
in the spinor multiplet in table~\ref{3mult1}. This and the higher
order terms in the eigenvalue will eventually be compared with 
the dimensions of operators in the closed, fermionic $\su(2|3)$ sector in the 
gauge theory. 

The argument we are making relies heavily on the claim that the perturbation
matrix is block diagonal on the closed subsectors described above: we have evaluated
the exact energy shift on these subsectors by simply taking the diagonal
matrix element of the perturbing Hamiltonian in a particular state in each
sector. We will now carry out a simple numerical test of the claimed
block diagonalization of the full perturbing Hamiltonian. The basic idea
is that, while it is impractical to algebraically diagonalize the full
$2,048\times 2,048$ perturbation matrices, it is quite easy to do a numerical
diagonalization for a specific choice of $\lambda'$ and mode indices 
$q,r,s$. One can then check that the numerical eigenvalues match
the analytic predictions evaluated at the chosen coupling and mode indices. 
For definiteness, we choose
\be
q = 1 \qquad r=2 \qquad s=-3 \qquad \lambda' = 1\ .
\label{MIvals}
\ee
The predicted eigenvalue shifts of the three protected states, evaluated
at the parameter choices of (\ref{MIvals}) are given in table~\ref{centroidnum}.
These values come directly from eqns.~(\ref{exactSO41},\ref{exactSO42},\ref{exactfermi})
above (with $J$ set to unity, for convenience).
\begin{table}[ht!]
\begin{eqnarray}
\begin{array}{|ccc|}
\hline
	\delta E: ~\lambda' = 1 & & q=1,\ r=2,\ s=-3 \\ 
\hline
 	\delta E_{AdS}(1,2,-3,J=1) &=& -16.255434067000426 \\ 
  	\delta E_{S^5}(1,2,-3,J=1) &=&  -20.137332508389193 \\ 
	\delta E_{\rm Fermi}(1,2,-3,J=1) &=& -18.19638328769481 \\
\hline
\end{array} \nonumber
\end{eqnarray}
\caption{Exact numerical eigenvalues of three-impurity protected sectors}
\label{centroidnum}
\end{table}
Since we want to compare these energies to a numerical diagonalization, 
we must maintain a high level of precision in the numerical computation. 
With the parameter choices of (\ref{MIvals}), the numerical diagonalization
of the full $2,048\times 2,048$ perturbation matrices 
on both the spacetime boson (table~\ref{boseblock3}) and spacetime 
fermion (table~\ref{fermiblock3}) sectors yields the spectrum and multiplicities
displayed in table~\ref{numfermi}. The multiplicities are consistent with the
superconformal multiplet structure we found in 
the one-loop analysis (given in table~\ref{3mult1}). The predicted 
closed sector eigenvalues (listed in table~\ref{centroidnum}) match, to the 
precision of the calculation, entries in the list of numerical eigenvalues. 
These energies also appear at the expected levels within the multiplets.
$E_{AdS}(1,2,-3,J)$ and $E_{S^5}(1,2,-3,J)$ appear in bosonic levels with 
multiplicity $280_B$, while energy $E_{\rm Fermi}(1,2,-3,J)$ appears as a 
fermionic level with multiplicity $560_F$; according to table~\ref{3mult1}
these are uniquely identified as the central $L=4$ levels of their respective
multiplets, exactly where the protected energy levels must lie. All of 
this is clear evidence that the `closed sector' states of the string theory 
do not mix with other states under the perturbing Hamiltonian, thus justifying
our method of calculating their exact eigenenergies.
\begin{table}[ht!]
\begin{eqnarray}
\begin{array}{|cc|}
\hline
\delta E(1,2,-3,J=1)\ \lambda' = 1 & {\rm Mult.}  \\
\hline
          -30.821354623065    &                     4_B\\
         -26.9394561816763    &                     4_B\\
         -26.2093998737015    &                    64_B\\
         -25.4793435657269    &                   112_B\\
         -21.5974451243382    &                   112_B\\
         -20.8673888163637    &                   448_B\\
         -20.1373325083891    &                   280_B\\
         -16.2554340670003    &                   280_B\\
         -15.5253777590258    &                   448_B\\
         -14.7953214510512    &                   112_B\\
         -10.9134230096624    &                   112_B\\
         -10.1833667016878    &                    64_B\\
          -9.4533103937133    &                     4_B\\
         -5.57141195232456    &                     4_B\\
\hline
\end{array} \nonumber
\qquad
\begin{array}{|cc|}
\hline
\delta	E(1,2,-3,J=1)\ \lambda' = 1 & {\rm Mult.}  \\
\hline
         -28.8804054023706    &                         8_F\\
          -28.150349094396    &                        32_F\\
         -24.2684506530072     &                       32_F\\
         -23.5383943450326     &                      224_F\\
          -22.808338037058     &                      224_F\\
         -18.9264395956693     &                      224_F\\
         -18.1963832876947    &                       560_F\\
         -17.4663269797201    &                       224_F\\
         -13.5844285383314    &                       224_F\\
         -12.8543722303568    &                       224_F\\
         -12.1243159223822    &                        32_F\\
         -8.24241748099347    &                        32_F\\
         -7.51236117301893    &                         8_F\\
\hline
\end{array} \nonumber\\
\end{eqnarray}
\caption{All loop numerical spectrum of three-impurity states ($q=1,~r=2,~s=-3,~
\lambda'=1,~J=1$). Left panel: bosons; right panel: fermions }
\label{numfermi}
\end{table}


At one loop, we found that the three superconformal multiplets were
displaced from each other by precisely the internal level spacing.
This led to an accidental degeneracy which is lifted in the exact 
dimension formulas we have just derived. To explore this, it is useful
to have formulas for the eigenvalues of all the levels in each multiplet. 
>From the discussion in section 3.3, we see that
each level in the string energy spectrum can be connected by
a simple integer shift in the angular momentum $J$.
Since we are working at $O(1/J)$ in a large-$J$ expansion, 
all contributions from this shift must come from the BMN
limit of the theory.  In other words, by sending
$J\to J+2-L/2$ in the BMN formula for the energy
\be
E = \sqrt{1+\frac{n^2 g_{YM}^2 N_c}{(J+2-L/2)^2}} + \dots\ ,	
\ee
we can generate an expansion, to arbitrary order in $\lambda'$,
for each level $L$ in the entire superconformal multiplet. 

For the vector $SO(4)_{AdS}$ multiplet, we find 
\be
\delta E_{AdS}(q,r,J,L) & \approx & 
	\frac{1}{J}\biggl\{
	(L-6)(q^2+qr+r^2)\lambda'
	-\frac{1}{2}(L-4)(q^2+qr+r^2)^2	{\lambda'}^2
\nn\\
&&	+\frac{3}{16}\Bigl[
	2(L-4)q^6
	+6(L-4)q^5r
	+5(3L-14) q^4r^2
	+20(L-5) q^3r^3
\nn\\
&&\kern+35pt	+5(3L-14) q^2r^4
	+6(L-4) qr^5
	+2(L-4)r^6
	\Bigr]{\lambda'}^3
\nn\\
&&\kern-75pt	-\frac{(q^2+qr+r^2)}{16}\Bigl[
	5(L-4)q^6
	+15(L-4)q^5r
	+(50L-247)q^4r^2
	+(75L-394)q^3r^3
\nn\\
&&\kern+0pt	+(50L-247)q^2r^4
	+15(L-4)qr^5
	+5(L-4)r^6
	\Bigr]{\lambda'}^4 + \dots
	\biggr\}\ 
\ee
(for convenience in eventual comparison with the gauge theory, the
eigenvalues have been expanded to $O(\lambda'^4)$). The corresponding
result for the $SO(4)_{S^5}$ vector multiplet is
\be
\delta E_{S^5}(q,r,J,L) & \approx & 
	\frac{1}{J}\biggl\{
	(L-8)(q^2+qr+r^2)\lambda'
	-\frac{1}{2}
	(L-6)(q^2+qr+r^2)^2
	{\lambda'}^2
\nn\\
&&	+\frac{3}{16}\Bigl[
	2(L-6)q^6
	+6(L-6)q^5r
	+(15L-92) q^4r^2
	+4(5L-31) q^3r^3
\nn\\
&&\kern+35pt	+(15L-92) q^2r^4
	+6(L-6) qr^5
	+2(L-6)r^6
	\Bigr]{\lambda'}^3
\nn\\
&&\kern-75pt	-\frac{(q^2+qr+r^2)}{16}\Bigl[
	5(L-6)q^6
	+15(L-6)q^5r
	+(50L-309)q^4r^2
	+3(25L-156)q^3r^3
\nn\\
&&\kern+00pt	+(50L-309)q^2r^4
	+15(L-6)qr^5
	+5(L-6)r^6
	\Bigr]{\lambda'}^4 + \dots
	\biggr\}\ .
\ee
Finally, the result for the spinor multiplet is
\be
\delta E_{\rm Fermi}(q,r,J,L) & \approx & 
	\frac{1}{J}\biggl\{
	(L-7)(q^2+qr+r^2)\lambda'
	-\frac{1}{2}
	(L-5)(q^2+qr+r^2)^2
	{\lambda'}^2
\nn\\
&&	+\frac{3}{16}\Bigl[
	2(L-5)q^6
	+6(L-5)q^5r
	+3(5L-27) q^4r^2
	+4(5L-28) q^3r^3
\nn\\
&&\kern+35pt	+3(5L-27) q^2r^4
	+6(L-5) qr^5
	+2(L-5)r^6
	\Bigr]{\lambda'}^3
\nn\\
&&\kern-75pt	-\frac{(q^2+qr+r^2)}{16}\Bigl[
	5(L-5)q^6
	+15(L-5)q^5r
	+2(25L-139)q^4r^2
	+(75L-431)q^3r^3
\nn\\
&&\kern+00pt	+2(25L-139)q^2r^4
	+15(L-5)qr^5
	+5(L-5)r^6
	\Bigr]{\lambda'}^4 + \dots
	\biggr\}\ .
\ee
It is important to remember that, to obtain the energies of the states as opposed
to the energy shifts $\delta E$, we must add the BMN energy of the original 
degenerate multiplet to the above results:
\be
\label{BMNen}
E_{\rm BMN} &=& \sqrt{1+\lambda' q^2}+\sqrt{1+\lambda' r^2}+\sqrt{1+\lambda' (q+r)^2}\nn\\
	&=& 3 + (q^2+r^2+qr)\lambda' - \frac{1}{4}(q^2+r^2+qr)^2\lambda'^2+ \ldots
\ee

We can conclude from the above formulas that all three multiplets have
a common internal level spacing given by the following function of 
$\lambda'$ and mode indices:
\be
\label{levelspacing}
\frac{\delta E}{\delta L}& \approx & 
	\frac{1}{J}\biggl\{
	(q^2+qr+r^2)\lambda'
	-\frac{1}{2}\Bigl[
	(q^2+qr+r^2)^2
	\Bigr]{\lambda'}^2
\nn\\
&&	+\frac{3}{16}\Bigl[2q^6 +6q^5r +15 q^4r^2 +20q^3r^3 
 +15q^2r^4 +6qr^5 +2r^6 \Bigr]{\lambda'}^3
\nn\\
&&\kern-25pt	-\frac{(q^2+qr+r^2)}{16}\Bigl[
	5q^6 +15q^5r +50q^4r^2 +75q^3r^3
	+50q^2r^4 +15qr^5 +5(r^6 \Bigr]{\lambda'}^4 + \dots
	\biggr\}\ .
\nn\\
&&
\ee
We have expanded in powers of $\lambda'$, 
but an all-orders formula can easily be constructed. The multiplets are 
displaced from one another by shifts that also depend on $\lambda'$
and mode indices. We note that the one-loop degeneracy between 
different multiplets (see table~\ref{3mult1}) is preserved to second 
order in $\lambda'$, but is broken explicitly at three loops. 
At this order and beyond, each multiplet acquires a constant overall 
($L$-independent) shift relative to the other two.

\subsection{Two equal mode indices: $(q = r = n,\ s=-2n)$}

An independent analysis is required when two mode indices are equal 
(specifically, we choose $q = r = n$, $s=-2n$). The all-loop matrix elements 
are complicated and we will refrain from giving explicit expressions for them
(though the complete formulas can be found at \cite{webref}). 
As in the unequal mode index case, however, exact eigenvalues can easily be
extracted by projection onto certain protected subsectors. In particular,
the energy shift for states created by three bosonic mode creation operators
with symmetric-traceless $SO(4)_{AdS}$ vector indices (the $\Sl(2)$ sector)
turns out to be
\be
\delta E_{AdS}(n,J)  & = &  
	-\frac{n^2\lambda'}{J(1+n^2\lambda')\sqrt{4n^2+1/\lambda'}}
	\biggl\{
	\sqrt{4n^2+\frac{1}{\lambda'}}\left( 3+4n^2\lambda'\right)
	+\omega_n\left( 4+8n^2\lambda'\right)
	\biggr\}
\nn\\
	& \approx & 
	\frac{1}{J}\biggl\{
	-7n^2\lambda' + n^4{\lambda'}^2 - \frac{17}{2}n^6{\lambda'}^3+\dots
	\biggr\}\ .
\label{nnval1}		
\ee
The leading order term in the small-$\lambda'$ expansion is the 
$-7/3~[280_B]$ level $L=4$ eigenvalue in the $\Lambda_2$ multiplet
in table~\ref{nnspec}. The energy shift of the $SO(4)_{S^5}$ partners of 
these states (belonging to the $\su(2)$ closed sector) is
\be
\delta E_{S^5}(n,J)  & = &  
	-\frac{n^2\lambda'}{J(1+n^2\lambda')\sqrt{4n^2+1/\lambda'}}
	\biggl\{
	\sqrt{4n^2+\frac{1}{\lambda'}}\left( 5+4n^2\lambda'\right)
	+\omega_n\left( 6+8n^2\lambda'\right)
	\biggr\}
\nn\\
	& \approx & 
	\frac{1}{J}\biggl\{
	-11n^2\lambda' + 8n^4{\lambda'}^2 - \frac{101}{4}n^6{\lambda'}^3+\dots
	\biggr\}\ .		
\label{nnval2}
\ee
The one-loop correction corresponds to the $-11/3~[280_B]$ level
in the $\Lambda_1$ submultiplet of table~\ref{nnspec}.
As noted above, the protected symmetrized-fermion ($\su(2|3)$) sector does not
appear when two mode indices are equal. As in the previous section,
we can do a numerical diagonalization of the full perturbation matrix 
to verify that the predicted eigenvalues are indeed
exact and closed, but we will omit the details.

By invoking the angular momentum shift $J\to J+2-L/2$ in the BMN limit, 
we can use the energy shift of the $L=4$
level to recover the exact energy shifts of all other levels
in the superconformal multiplets of table~\ref{nnspec}. The energy shifts of the 
vector multiplet containing the protected $SO(4)_{AdS}$ bosonic irrep 
at level $L=4$ are given by the expression
\be
\delta E_{AdS}(n,J,L) & \approx & 
	\frac{1}{J}\biggl\{
	\frac{1}{2}(3L-19)n^2{\lambda'}
	-\frac{1}{2}(9L-38)n^4{\lambda'}^2
	+\frac{1}{8}(99L-464)n^6{\lambda'}^3
\nn\\
&&\kern+100pt	-\frac{1}{16}(645L-3160)n^8{\lambda'}^4
	+\dots
	\biggr\}\ .
\ee
The shifts of the multiplet containing the protected $SO(4)_{S^5}$ 
bosonic irrep are given by
\be
\delta E_{S^5}(n,J,L) & \approx & 
	\frac{1}{J}\biggl\{
	\frac{1}{2}(3L-23)n^2{\lambda'}
	-\frac{1}{2}(9L-52)n^4{\lambda'}^2	
	+\frac{1}{8}(99L-598)n^6{\lambda'}^3
\nn\\
&&\kern+100pt
	-\frac{1}{16}(645L-3962)n^8{\lambda'}^4
	+\dots
	\biggr\}\ .
\ee
Once again, we note that in order to get energies, rather than energy shifts, 
one must append the BMN energy of the original degenerate multiplet to these 
results. Unlike the unequal mode index case, there is no accidental degeneracy
between superconformal multiplets spanning the three-impurity
space, even at one loop in $\lambda'$. The level spacings within the two 
superconformal multiplets are the same, but the multiplets are offset from each 
other by an $L$-independent shift (but one that depends on $\lambda'$ and
mode indices).  

\section{Gauge theory anomalous dimension comparison}
In the previous sections, we have given a complete analysis of the perturbed
energy spectrum of three-impurity string states. The `data' are internally
consistent in the sense that the perturbed energy levels organize themselves 
into proper superconformal multiplets of the classical 
nonlinear sigma model governing the string worldsheet dynamics. Since the 
quantization procedure leaves only a subgroup of the full symmetry group 
as a manifest, linearly realized symmetry, this is by itself a nontrivial 
check on the consistency of the action and quantization procedure. To address 
the issue of AdS/CFT duality, we must go further and compare the string 
energy spectrum with the anomalous dimensions of gauge theory operators 
dual to the three-impurity string states. 

The task of finding the anomalous dimensions of BMN operators in the limit
of large $R$-charge and dimension $D$, but finite $\Delta=D-R$,
is greatly simplified by the existence of an equivalence between 
the dilatation operator of ${\cal N}=4$ SYM and the Hamiltonian of a 
one-dimensional spin chain. This correspondence was first
proposed by Minahan and Zarembo \cite{Minahan:2002ve}, who showed that
the anomalous dimension matrix in the $\so(6)$ subsector of the theory, 
expanded to one loop in the 't Hooft coupling $\lambda = g_{YM}^2 N_c$, 
is equivalent to the Hamiltonian of an integrable spin chain. 
Beisert and Staudacher then showed that a more elaborate integrable spin chain 
describes the action of the one-loop anomalous dimension operator on the general 
single-trace operator \cite{Beisert:2003yb}. 
(The complete one-loop dilatation operator was derived in \cite{Beisert:2003jj}.)
The key to this development is the fact that the one-loop spin chain Hamiltonian
has only nearest-neighbor interactions (in the planar large-$N_c$ limit)
and is of limited complexity.
Higher-loop gauge theory physics is encoded in increasingly long-range 
spin chain interactions which generate a rapidly growing number 
of possible terms in the Hamiltonian \cite{Beisert:2003tq}.  
Fixing the coefficients of all
these terms by comparison with diagrammatic computations would be a very
impractical approach.  Fortunately, Beisert was able to show that, at 
least for BMN operators in the $\su(2)$ closed subsector, 
general requirements (such as the existence of a well-defined BMN scaling limit) 
suffice to fix the form of the spin chain 
Hamiltonian out to three-loop order \cite{Beisert:2003ys,Beisert:2003jb}. 
In this section we will
discuss the use of these higher-loop spin chains to generate the information
we need on the anomalous dimensions of three-impurity operators. 
We will summarize the salient points here, leaving most of the details
to a separate paper \cite{spinchain}.

We have already noted that there are three closed subsectors of BMN 
operators in which impurities taken from a subalgebra of the full 
superconformal algebra mix only with themselves: we have referred to them
as the $\Sl(2)$, $\su(2)$ (both bosonic) and $\su(2|3)$ (fermionic) sectors. We will focus
our attention on these sectors because their spin chain description is
simple and their anomalous dimensions fix the dimensions of the 
remaining three-impurity operators in the theory.  Spin-chain Hamiltonians incorporating
higher-loop-order gauge theory physics have been constructed for the
$\su(2)$ and $\su(2|3)$ sectors but, as far as we know, the $\Sl(2)$ spin
chain is known only to one-loop order. 

Although these spin chains are integrable, methods such as the Bethe ansatz 
technique do not immediately yield the desired results for multiple-impurity anomalous 
dimensions. Minahan and Zarembo did use the Bethe ansatz for the one-loop $\so(6)$ 
spin chain (of which the exactly closed $\su(2)$ system is a subsector) to obtain 
approximate multi-impurity anomalous dimensions \cite{Minahan:2002ve}, but we 
need results for all sectors and for higher-loop spin chains. As mentioned above,
the $\Sl(2)$ spin chain has phenomenological applications and has been extensively
developed in that context. It is therefore possible that some of the results we need 
can be extracted from the relevant literature.\footnote{We thank A.~Belitsky for
making us aware of this literature and for helpful discussions on this point.}
In the end, since we are looking for a unified approach that can handle all 
sectors and any number of loops, we decided that numerical methods are, at the
moment, the most effective way to extract the information we need about gauge 
theory anomalous dimensions.  Since Bethe ansatz equations exist for most of the
results that are of interest to us, the numerical results obtained here can 
eventually be checked against the Bethe-ansatz methodology:  these exercises
will be reserved for \cite{spinchain}.

We begin with a discussion of the bosonic $\Sl(2)$ sector.
For total $R$-charge ${\alg L}$ (the $R$-charge is equal to the number of lattice
sites ${\alg L}$ in this sector), the basis for this system consists of 
single-trace operators of the form
\be
{\rm Tr}\left({\cal D}^IZ~Z^{{\alg L}	-1} \right), 
~{\rm Tr}\left({\cal D}^{I-1}Z~{\cal D}Z~Z^{{\alg L}	-2} \right),
	~{\rm Tr}\left({\cal D}^{I-1}Z~Z{\cal D}Z~Z^{{\alg L}	-3} \right),~ \ldots~,
\ee
where $Z$ is the $SO(6)$ Yang-Mills boson carrying one unit of $R$-charge,
${\cal D}$ is a spacetime covariant derivative operator that scales under
the chosen $\Sl(2)$ subgroup of the Lorentz group, $I$ is the total impurity
number and the full basis contains all possible distributions of ${\cal D}$
operators among the $Z$ fields. Conservation of various $U(1)$ subgroups
of the $R$-symmetry group ensures that operators of this type mix only among
themselves to all orders in the gauge theory (as long as we work in the 't Hooft
large-$N_c$ limit). This gauge theory closed subsector corresponds to the symmetric 
traceless irrep of $SO(4)_{AdS}$ bosons in the string theory (states whose
energy shifts are given in eqns.~(\ref{exactSO41}) and (\ref{nnval1})). 

The one-loop spin chain Hamiltonian for this sector has been derived by Beisert 
\cite{Beisert:2003jj} in a representation where there is a lattice site assigned to
each $Z$ field, and each site supports a harmonic oscillator whose level of
excitation counts the number of ${\cal D}$ operators acting on that $Z$
insertion.  The raising operator $a_i^\dag$ therefore corresponds to the 
insertion of a derivative at the $i^{\rm th}$ lattice site:
\be
\ket{\alg L} \sim {\rm Tr} \left(Z^{\alg L} \right)\ ,\qquad 
(a_i^\dag)^n \ket{\alg L} \sim {\rm Tr} \left( Z^{i-1} {\cal D}^n Z  Z^{{\alg L}-i} \right)\ ,
\ \ldots 
\ee
The spin chain Hamiltonian is a sum over $\alg L$ lattice sites of a nearest-neighbor 
interaction which moves excitations between neighbors while keeping 
the net excitation number (number of impurities $I$) fixed:
\be
&&H^{\Sl(2)}  = 
	\frac{{\lambda}}{{8\pi^2}}\sum_{k=1}^{\alg L} H^{\Sl(2)}_{k,k+1}\ , 
\nn\\
&&H^{\Sl(2)}_{1,2} (a_1^\dag)^k (a_2^\dag)^{n-k}\ket{\alg L}  =  
	\sum_{k'=0}^{n}\left[
	\delta_{k=k'}\left(h(k)+h(n-k)\right)
	-\frac{\delta_{k\neq k'}}{|k-k'|}\right]
	(a_1^\dag)^{k'}(a_2^\dag)^{n-k'}\ket{{\alg L}}\ .
\nn\\
&&
\label{sl2int}
\ee
The weighting of different terms by the harmonic numbers $h(n) = \sum_{r=1}^n r^{-1}$
is a general feature of spin chains derived from gauge theories and reflects the
infrared behavior of gluon emission.\footnote{We are indebted to A.~Belitsky for
clarifying discussions on this point.} 
The impurity number $I$ is the
net excitation number of the $\alg L$ oscillators. The interaction propagates impurities 
around the lattice and assigns special amplitudes to `collisions' of multiple
excitations on a single site. Our interest is in finding the spectrum of energies of
three impurities on a large lattice under these dynamics. 

When $\alg L$ is large and the number of impurities is fixed, it is natural to
seek a virial expansion for the eigenvalues of $H^{\Sl(2)}$. The above Hamiltonian 
acts on an isolated impurity exactly like the usual lattice Laplacian; it 
can be diagonalized by passing to momentum space. The rest of the Hamiltonian
amounts to two- and higher-body  scattering vertices for the single-impurity 
pseudoparticles, and standard many-body intuition tells us that such interaction 
vertices are suppressed by powers of $\alg L^{-1}$ compared to the leading pseudoparticle 
energies. Since we are in fact looking for an expansion of the energy eigenvalues 
in powers of $\alg L^{-1}$, it is useful to rewrite $H^{\Sl(2)}$ in terms of momentum
space creation and annihilation operators $a_p,~ a^{\dag}_p$ 
defined by the usual discrete Fourier transform.  Since the definition of $H^{\Sl(2)}$
is given in terms of its action on states, rather than as an explicit operator,
this rewriting of the Hamiltonian takes a little work, but the result is simple
\cite{heckman}:
\begin{eqnarray}
&&	H^{\Sl(2)}  = \frac{\lambda}{8\pi^2} \Bigl[
	~\underset{p}{\sum}4\sin^{2}\frac{p \pi}{\alg L}~a_{p}^{\dag}a_{p} \nn\\
&&\kern+25pt 	+ \frac{1}{\alg L} \underset{p,q,r,s}{\sum}\delta_{p+q,r+s}
		\left(-\sin^{2}\frac{p\pi}{\alg L}
		-\sin^{2}\frac{q\pi}{\alg L}+\sin^{2}\frac{(p+q)\pi}{\alg L}\right)  
		a_{p}^{\dag}a_{q}^{\dag }a_{r}a_{s}~ \Bigr]
		+\ldots
\label{hamilsltwofin}
\end{eqnarray}
The pseudoparticle creation and annihilation operators $a_p,~ a^{\dag}_p$ are labeled
by integer momenta $p=0,1,\ldots,\alg L-1$ and obey the standard algebra. The ellipses 
stand for three- and higher-body interactions, which are suppressed by even higher 
powers of $\alg L^{-1}$. For $I$ impurities 
carrying quantized pseudoparticle momenta $n_i$, this leads to an energy formula
\be
\label{pseudoen}
E_{\alg L} =I + \frac{\lambda}{2 \pi^2}\underset{i=1}{\overset{I}{\sum}}
	\sin^2\frac{n_{i}\pi}{\alg L} + \frac{\lambda}{\alg L^3} 
	V_{\rm two-body}(n_1,\ldots,n_I) + \ldots
\ee
To facilitate comparison with string theory, we have reinstated the zeroth-order term  
for the total dimension minus $R$-charge of an $I$-impurity operator. The true 
eigenvalues differ by small corrections from the lattice Laplacian energies of the 
free pseudoparticles, which are labeled by the pseudoparticle momenta $n_i$. 
These integers are the gauge theory analogs of the string mode indices and, to make 
contact with string theory, we must take the limit $\alg L\to\infty$ while keeping the 
$n_i$ fixed. In this limit, the eigenvalues of eqn.~(\ref{pseudoen}) will scale as
\be
\label{scaling}
E_{\alg L}(\lbrace n_i \rbrace) = I + \frac{\lambda}{\alg L^2}{E^{(1,2)}
	(\lbrace n_i \rbrace)} + 
\frac{\lambda}{\alg L^3}{E^{(1,3)}(\lbrace n_i \rbrace)} + O(\lambda \alg L^{-4})\ .
\label{expG1}
\ee
This is the leading-order (in $\lambda$) version of the more general BMN scaling
\be
\label{BMNscale}
E(\lambda,J)\simeq E(\lambda/J^2,1/J) = 
	\sum_{i=1,\ j=0 } E^{(i,2i+j)}\left(\frac{\lambda}{J^2}\right)^i \ J^{-j} \ ,
\ee
which, as we have seen in previous sections, naturally characterizes
string energy levels. In the case at hand, where we are computing energies to 
$O(\lambda)$ only, BMN scaling is a consequence of the $\alg L^{-2}$ scaling of energy
eigenvalues that follows automatically from the form of the virial Hamiltonian 
(\ref{hamilsltwofin}).  Whether the spectra of higher-order spin chain Hamiltonians 
scale with $\alg L$ in accordance with eqn.~(\ref{BMNscale}) is a very nontrivial question
which we will address shortly.

To compare with the corresponding string theory predictions of eqns.~(\ref{exactSO41}) 
and (\ref{nnval1}), we reorganize those results as follows: we reinstate the 
BMN energy of the degenerate multiplet (\ref{BMNen}) (expanded to first order in 
$\lambda'$); we replace $\lambda'$ with $\lambda/J^2$ and replace $J$ by $\alg L$. 
This gives specific string theory predictions for the large-$\alg L$ scaling of one-loop
anomalous dimensions of the AdS, or $\Sl(2)$, closed sector. As usual, there are
two distinct cases: for unequal mode indices $(q\ne r\ne s=-q-r)$, we have
\be
E_{AdS}(q,r,\alg L) = 3 + (\alg L-2)(q^2 + r^2 + qr)\frac{\lambda}{\alg L^3} 
		+ O(\alg L^{-4})\ ,
\ee 
while for pairwise equal mode indices $(n,n,-2n)$ we have
\be
E_{AdS}(n,\alg L) = 3 + (3 \alg L-7)n^2\frac{\lambda}{\alg L^3} + O(\alg L^{-4})\ .
\ee
This matches the expected virial scaling of the spin chain eigenvalues 
displayed in eqn.~(\ref{scaling}), with the specific identifications
\be
E_{AdS}^{(1,2)} = (q^2 + r^2 + qr) \qquad E_{AdS}^{(1,3)} = -2(q^2 + r^2 + qr) \qquad
E_{AdS}^{(1,3)}/E_{AdS}^{(1,2)} = -2
\label{expSL21}
\ee
for $q\neq r\neq s=-q-r$, or
\be
E_{AdS}^{(1,2)} = 3 n^2 \qquad E_{AdS}^{(1,3)} = -7 n^2  \qquad
E_{AdS}^{(1,3)}/E_{AdS}^{(1,2)} = -7/3
\label{expSL22}
\ee
for $q=r=n$ and $s=-2n$.

To check these predictions, we numerically diagonalize the spin chain for 
three impurities for a sequence of values of $\alg L$ 
(up to $30 \lesssim \alg L \lesssim 60$ in practice). 
It is a matter of convenience whether we construct the $\alg L\times \alg L$ 
Hamiltonian matrix 
using the position space version (\ref{sl2int}) or the momentum space version
(\ref{hamilsltwofin}) of the Hamiltonian. We then track how the eigenvalues 
evolve as $\alg L$ varies and fit the data to a general $\alg L^{-1}$ expansion in order to
extract the spin chain coefficients 
$E_{\Sl(2)}^{(1,2)},\ E_{\Sl(2)}^{(1,3)}$ (the vanishing of the coefficient
$E_{\Sl(2)}^{(1,1)}$ is a check of BMN scaling, but is essentially guaranteed here). The 
results of this exercise, presented in table~\ref{NUM_SL(2)_1Loop}, show 
clear agreement between the string and gauge theory predictions. 
In the fourth column we list the ladder of string mode 
indices that correspond to the succession of eigenstates in the gauge theory, 
according to eqns.~(\ref{expSL21}) and (\ref{expSL22}). For the low-lying states in the 
spectrum, the numerical agreement is convincing.  As one moves up the ladder of energies, 
higher-order $1/\alg L$ 
corrections become more important. We would have to generate data up to higher lattice 
sizes and do a more precise scaling fit to improve the agreement at higher levels
in the spectrum. 
\begin{table}[ht!]
\begin{eqnarray}
\begin{array}{|ccccc|}
\hline
E_{\Sl(2)}^{(1,2)} & E_{\Sl(2)}^{(1,3)} & E_{\Sl(2)}^{(1,3)}/E_{\Sl(2)}^{(1,2)}  
                                  & {\rm String\ Modes\ }(q,r,s)
	& {\rm \%\ Error}  \\
\hline
1+1.2\times 10^{-9}	&-2-3.1\times 10^{-7}	&-2-3.1\times 10^{-7}	& (1,0,-1)	&0.00002	\% \\
3-7.6\times 10^{-9}	&-7+1.9\times 10^{-6}   &-7/3+6.3\times 10^{-7} & (1,1,-2)	&0.00001	\% \\
3-7.6\times 10^{-9}	&-7+1.9\times 10^{-6}	&-7/3+6.3\times 10^{-7}	&  (-1,-1,2)	&0.00001	\% \\
4-2.8\times 10^{-7}	&-8+6.9\times 10^{-6}   &-2+1.7\times 10^{-6}   &  (2,0,-2)	&0.0001	\% \\
7-2.9\times 10^{-7}	&-14+7.1\times 10^{-5}	&-2+1.0\times 10^{-5}	&  (1,2,-3) 	&0.0005	\% \\
7-2.9\times 10^{-7}	&-14+7.1\times 10^{-5}	&-2+1.0\times 10^{-5}	&  (-1,-2,3)	&0.0005	\% \\
9-4.1\times 10^{-7}	&-18+1.0\times 10^{-4}	&-2+1.0\times 10^{-5}	&  (3,0,-3)	&0.0005	\% \\
12+8.4\times 10^{-7}	&-28-1.5\times 10^{-4}	&-7/3-1.2\times 10^{-5}	&  (2,2,-4)	&0.0003	\% \\
12+8.4\times 10^{-7} 	&-28-1.5\times 10^{-4}	&-7/3-1.2\times 10^{-5}	&  (-2,-2,4)	&0.0003	\% \\	
13-7.0\times 10^{-6}	&-26+1.7\times 10^{-3}	&-2+1.3\times 10^{-4}	&  (1,3,-4) 	&0.01	\% \\
13-7.0\times 10^{-6}	&-26+1.7\times 10^{-3}	&-2+1.3\times 10^{-4}	&  (-1,-3,4)	&0.01	\% \\
16-1.4\times 10^{-6}	&-32+3.9\times 10^{-4}	&-2+2.4\times 10^{-5}	&  (4,0,-4)	&0.002	\% \\
19-7.5\times 10^{-6}	&-38+2.2\times 10^{-3}	&-2+1.1\times 10^{-4}	&  (2,3,-5)	&0.01	\% \\
19-7.5\times 10^{-6}	&-38+2.2\times 10^{-3}	&-2+1.1\times 10^{-4}	&  (-2,-3,5)	&0.01	\% \\
21-3.4\times 10^{-6}	&-42+8.8\times 10^{-4}	&-2+4.2\times 10^{-5}	&  (1,4,-5)	&0.002	\% \\
21-3.4\times 10^{-6}	&-42+8.8\times 10^{-4}	&-2+4.2\times 10^{-5}	&  (-1,-4,5)	&0.002	\% \\
\hline
\end{array} \nonumber
\end{eqnarray}
\caption{Scaling limit of numerical spectrum of three-impurity
	$\Sl(2)$ sector at one-loop order}
\label{NUM_SL(2)_1Loop}
\end{table}


At this point it is appropriate to say a few words about the role of integrability
in this problem. It was first argued in \cite{Bena:2003wd} 
that the complete GS action of IIB superstring theory
on $AdS_5\times S^5$ is integrable.  Integrability has since taken a central role in
studies of the AdS/CFT correspondence, as any precise non-perturbative understanding
of integrability on both sides of the duality would be extremely powerful.
Integrability on either side of the duality gives rise to an infinite tower of
hidden charges that can be loosely classified as either local (Abelian) or non-local
(non-Abelian).
In the Abelian sector, contact between the integrable structures of 
gauge theory and semiclassical string theory (a subject which was first investigated in
\cite{0310182}) has been made to two loops in $\lambda$ 
(see, eg., \cite{Beisert:2003ea,0306139,0403077,0402207}).
(The corresponding problem in the non-local sector was addressed to one-loop 
order in \cite{Dolan:2003uh,Dolan:2004ps}.)
One of the local gauge theory charges, denoted by $Q_2$, can be shown to anticommute
in the $\su(2)$ sector with a parity operator $P$ (to three loops in $\lambda$), 
whose action on a single-trace state in the gauge theory
is to invert the order of all fields within the trace \cite{Beisert:2003ys,Beisert:2003tq}.
Furthermore, $Q_2$ can be shown to connect states of opposite parity.
Taken together with the conservation of $Q_2$, 
these facts imply that all eigenstates in the spectrum
connected by $P$ must be degenerate.  
These degenerate states are known as parity
pairs and their existence can be interpreted as a necessary 
(but not sufficient) condition for integrability. The spectrum in 
table~\ref{NUM_SL(2)_1Loop} exhibits such a degeneracy and makes it clear
that parity pairs are simply distinct states whose lattice momenta 
(or worldsheet mode indices)
are related by an overall sign flip.  Since the net momentum of allowed 
states is zero, parity pair states can in principle scatter into each other,
and their degeneracy is a non-trivial constraint on the interactions. 
As a small caveat, we note that lattice momentum conservation implies that
mixing of parity-pair states can only occur via connected three-body (or higher)
interactions.  As the virial analysis shows, at the order to which
we are working, only two-body interactions are present and the parity pair
degeneracy is automatic.  The same remark applies to the string theory analysis
to $O(J^{-1})$ in the curvature expansion.  A calculation of the string theory spectrum
carried out to $O(J^{-2})$ is needed to see whether parity pair degeneracy 
survives string worldsheet interactions; a discussion of this point will be
given in \cite{orderJ2}.


We now turn to the closed $\su(2)$ sector of gauge theory operators, 
corresponding to the symmetric-traceless bosonic $SO(4)_{S^5}$ 
sector of the string theory. The operator
basis for this sector consists of single-trace monomials built out of 
two complex scalar fields $Z$ and $\phi$, where $Z$ is the complex scalar 
carrying one unit of charge under the
$U(1)$ $R$-charge subgroup and $\phi$ is one of the two scalars with zero $R$-charge,  
transforming as an $SO(4)$ vector in the $SO(6)\simeq U(1)_R\times SO(4)$ decomposition 
of the full $R$-symmetry group of the gauge theory. The collection of operators
\be
\label{SO6basis}
\tr(\phi^I Z^{\alg L-I}),\ \tr(\phi^{I-1}Z\phi Z^{{\alg L-I}-1}),\ 
\tr(\phi^{I-2}Z\phi^2 Z^{{\alg L-I}-1}),\ \ldots
\ee
(and all possible permutations, modulo cyclic equivalence, of the ${\alg L}$ factors) forms 
a basis with $I$ impurities and $R$-charge equal to ${\alg L-I}$. The anomalous dimension
operator simply permutes these monomials among themselves in ways that get more 
elaborate as we go to higher loop orders in the gauge theory. An 
explicit spin-chain Hamiltonian which incorporates gauge theory physics up 
to three loops has been constructed by enumerating all possible interaction terms and 
fixing coefficients by demanding BMN scaling behavior of the spectrum for large
lattice size \cite{Beisert:2003ys,Beisert:2003jb,Beisert:2003tq}. We now turn to
a numerical analysis of scaling in this sector in order to examine the match to
string theory predictions at higher loop orders.

The complete Hamiltonian in this sector will be written as a sum of terms of
increasing order in the coupling constant $\lambda$:
\be
H^{\su(2)} = \sum_n \left( \frac{\lambda}{8 \pi^2} \right)^n H^{\su(2)}_{2n}\ .
\ee
The action of the different terms $H^{\su(2)}_{2n}$ 
on the operator basis of eqn.~(\ref{SO6basis}) can be built out of permutation operators 
$P_{ij}$ which exchange the fields on the $i^{\rm th}$ and $j^{\rm th}$ sites
on a lattice of $\alg L$ sites.  Using the compact notation 
\be
\{ n_1,n_2,\dots\} = \sum_{k=1}^L P_{k+n_1,\ k+n_1+1} P_{k+n_2,\ k+n_2+1}\dots\ ,
\ee
Beisert, Kristjansen and Staudacher 
\cite{Beisert:2003tq} find the following explicit forms for the one- and 
two-loop terms in the Hamiltonian (we will return to the question of three-loop
terms shortly):
\be
H^{\su(2)}_2 = 2\left( \{ \}-\{0\} \right)~, \qquad
	H^{\su(2)}_4 = 2\bigl( -4\{\} + 6\{0\} - (\{0,1\} + \{1,0\}) \bigr)\ .
\ee
Just as in the $\Sl(2)$ case, it is clear that for fixed impurity number $I$
and large lattice size $\alg L$, there is a virial expansion of the one-loop energy 
eigenvalues essentially identical to eqn.~(\ref{pseudoen}). We therefore expect 
the exact energy eigenvalues to be labeled by integer pseudoparticle momenta. 

To obtain the dependence of the spectrum on $\lambda$, our strategy will be to develop 
a standard Rayleigh-Schr\"odinger perturbation theory expansion, treating 
$H^{\su(2)}_2$ as a zeroth-order Hamiltonian (obtaining its eigenvalues 
and eigenvectors numerically), then using non-degenerate perturbation theory
in $H^{\su(2)}_4$ to obtain the next-order corrections (i.e.~taking expectation values
of $H^{\su(2)}_4$ in the eigenvectors of $H^{\su(2)}_2$). The expansion coefficients
of each eigenvalue are numbers which depend on the lattice size $\alg L$ in some 
non-explicit way: we have to do the calculation for many values of $\alg L$ and
perform an extrapolation in $\alg L^{-1}$ in order to find the information of
interest to us but, as we will see, this is not too difficult. The only possible 
obstruction to this program would be a degeneracy in the spectrum of $H^{\su(2)}_2$ 
which would oblige us to use degenerate perturbation theory. Although the spectrum of 
$H^{\su(2)}_2$ is indeed degenerate, the higher charge $Q_2$ constrains matrix 
elements of the perturbing Hamiltonian $H^{\su(2)}_4$ in such a way that a) the 
first-order perturbation theory calculation can proceed as if the spectrum were 
non-degenerate and b) the degeneracy is maintained to this order. 


These considerations lead us to a numerical scheme involving a series of steps. 
First, we find the eigenvalues and eigenvectors of $H^{\su(2)}_2$ for the three-impurity 
Hamiltonian on lattices of length up to $\alg L\approx 30$ and fit a power 
series in $\alg L^{-1}$ to the eigenvalues.  Table~\ref{NUM_SO(6)_1Loop} displays the values 
of the coefficients $E_{\su(2)}^{(1,2)}, E_{\su(2)}^{(1,3)}$ (as defined in eqn.~\ref{expG1}) 
for the low-lying levels that we infer from this fit.  Second, we obtain a series of 
values for the $O(\lambda^2)$ corrections to the 
eigenvalues by taking the expectation value of the perturbing Hamiltonian $H^{\su(2)}_4$
between the numerical eigenvectors obtained in the previous step.
We fit a power series in $\alg L^{-1}$ to this data to read off the expansion
coefficients $E_{\su(2)}^{(2,n)}$ for the low-lying levels,
with the results displayed in table~\ref{NUM_SO(6)_2Loop}. It is important that the 
$O(\lambda^2)$ data scales as $\alg L^{-4}$ (i.e.~that the coefficients of lower powers
of $\alg L^{-1}$ vanish to numerical accuracy) as required by BMN scaling (see 
eqn.~(\ref{BMNscale})).  Since the Hamiltonian was determined in part by requiring this 
scaling, this is perhaps not a surprise: the real test will be the match of the BMN scaling 
coefficients to string theory data. 


To compare with string theory results for the bosonic symmetric-traceless
$SO(4)_{S^5}$ sector eigenvalues, we need to recast eqns.~(\ref{exactSO42}) 
and (\ref{nnval2}) as expansions in powers of $\lambda$ and $\alg L^{-1}$. 
We denote by $E_{S^5}^{(n,m)}$ the coefficient of $\lambda^n \alg L^{-m}$ 
in the large-$\alg L$ expansion of the string theory energies: they can be
directly compared with the corresponding quantities extracted from 
the numerical spin chain analysis. The string theory 
predictions for scaling coefficients, up to second order in $\lambda$, are
given in table~\ref{strscale}.
\begin{table}[ht!]
\begin{eqnarray}
\begin{array}{|c|cc|}
\hline
E_{S^5}^{(n,m)} & (q\ne r\ne s)  &  (q=r=n) \\
\hline  
E_{S^5}^{(1,2)} & (q^2 + qr + r^2) & 3n^2 \\
E_{S^5}^{(1,3)} & 2(q^2 + qr + r^2) & 7n^2 \\
E_{S^5}^{(2,4)} & -\frac{1}{4}(q^2 + qr + r^2)^2 & -\frac{9}{4} n^4 \\
E_{S^5}^{(2,5)} & -2(q^2 + qr + r^2)^2 & -19n^4 \\
\hline
\end{array} \nonumber
\end{eqnarray}
\caption{String predictions for $\su(2)$ scaling coefficients, to two loops}
\label{strscale}
\end{table}
As usual, the predictions for three-impurity states with unequal mode indices have
to be stated separately from those for states with two equal mode indices. The level
of agreement between string theory predictions and the spin chain numerical results is 
stated in the last column of tables~\ref{NUM_SO(6)_1Loop} and \ref{NUM_SO(6)_2Loop}.
The fit is very good for low-lying levels and gets worse as we go up in the spectrum. 
We are confident that, at any given level, the fit may be
made arbitrarily precise by generating data out to large enough lattice size.
We take these results as strong evidence that the string theory analysis agrees 
with the gauge theory up to $O(\lambda^2)$ in this sector. 
\begin{table}[ht!]
\begin{eqnarray}
\begin{array}{|ccccc|}
\hline
E_{\su(2)}^{(1,2)} & E_{\su(2)}^{(1,3)} & E_{\su(2)}^{(1,3)}/E_{\su(2)}^{(1,2)}  
& {\rm String\ Modes\ }(q,r,s) &
	{\rm \%\ Error}  \\
\hline
1+1.1\times 10^{-9}	& 2-1.6\times10^{-7}	&	2-1.7\times 10^{-7}	& (1,0,-1)	& 0.00001\% \\   
3-9.9\times 10^{-9}	& 7+1.5\times 10^{-6}	&	7/3+5.2\times 10^{-7}	& (1,1,-2)	& 0.00002\%\\
3-1.2\times 10^{-8}	& 7+1.8\times 10^{-6}	&	7/3+6.2\times 10^{-7}	&  (-1,-1,2)	& 0.00003\% \\
4-9.7\times 10^{-8}	& 8+1.5\times 10^{-5}	&	2+3.8\times 10^{-6}	&  (2,0,-2)	& 0.0002\% \\
7-1.6\times 10^{-6}	& 14+2.6\times 10^{-4}	&	2+3.8\times10^{-5}	&  (1,2,-3)	& 0.002\% \\
7-1.7\times 10^{-6}	& 14+2.6\times 10^{-4}	&	2+3.8\times10^{-5}	&  (-1,-2,3)	& 0.002\% \\
9-2.7\times 10^{-6}	& 18+4.9\times 10^{-4}	&	2+5.5\times10^{-5}	&  (3,0,-3)	& 0.003\%\\
12+3.1\times10^{-3}		& 27.5		&	2.29			&  (2,2,-4)	& 2\%\\
12-1.1\times 10^{-2}		& 29.9		&	2.50			&  (-2,-2,4)	& 7\%\\	
\hline
\end{array} \nonumber
\end{eqnarray}
\caption{Scaling limit of one-loop numerical spectrum of three-impurity 
	$\su(2)$ subsector}
\label{NUM_SO(6)_1Loop}
\end{table}
\begin{table}[ht!]
\begin{eqnarray}
\begin{array}{|ccccc|}
\hline
E_{\su(2)}^{(2,4)} & E_{\su(2)}^{(2,5)} & E_{\su(2)}^{(2,5)}/E_{\su(2)}^{(2,4)}  
& {\rm String\ Modes\ }(q,r,s) &
	{\rm \%\ Error}  \\
\hline
-0.24999997	& -2.000004	&	8.00002		& (1,0,-1)	& 0.0003\% \\   
-2.24999	& -19.002	&	8.445		& (1,1,-2)	& 0.01\%\\
-2.24999	& -19.002	&	8.445		&  (-1,-1,2)	& 0.01\% \\
-4.00003	& -31.997	&	7.9991		&  (2,0,-2)	& 0.01\% \\
-12.25		& -97.99	&	7.999		&  (1,2,-3)	& 0.02\% \\
-12.25		& -97.99	&	7.999		&  (-1,-2,3)	& 0.02\% \\
-20.25		& -161.8	&	7.990		&  (3,0,-3)	& 0.1\%\\
-36.10		& -289.6	&	8.02		&  (2,2,-4)	& 5.0\%\\
-36.18		& -276.4	&	7.64		&  (-2,-2,4)	& 9.5\%\\	
\hline
\end{array} \nonumber
\end{eqnarray}
\caption{Scaling limit of two-loop numerical spectrum of three-impurity 
	$\su(2)$ subsector}
\label{NUM_SO(6)_2Loop}
\end{table}

We now turn to a discussion of gauge theory physics beyond two loops. As it happens,
the three-loop Hamiltonian can be fixed up to two unknown coefficients 
($\alpha_1$ and $\alpha_2$) by basic field theory considerations \cite{Beisert:2003jb}:
\be
H^{\so(6)}_6 & = & 
	\left(60 + 6\alpha_1 -56\alpha_2\right)\{\}
	+ \left(-104 + 14\alpha_1 +96\alpha_2\right)\{0\}
\nn\\
&&	+ \left(24 + 2\alpha_1 -24\alpha_2\right)\left( \{0,1\}+\{1,0\} \right)
	+ \left(4 + 6\alpha_1 \right) \{0,2\}
\nn\\
&&	\left(-4 + 4\alpha_2 \right)\left( \{0,1,2\}+\{2,1,0\} \right)
	- \alpha_1\left( \{0,2,1\}+\{1,0,2\} \right)\ .
\label{H3SO6}
\ee 
Originally, these coefficients were determined by demanding proper BMN scaling 
in the theory and that the dynamics be integrable at three loops 
(by requiring that the Hamiltonian $H$ commute
with $Q_2$ up to three loops, and that $Q_2$ anticommute with $P$);
these assumptions set $\alpha_{1,2} = 0$.  By studying an $\su(2|3)$ 
spin chain model, Beisert \cite{Beisert:2003ys} was subsequently able to show that 
independent symmetry arguments, along with BMN scaling,
uniquely set $\alpha_1 = \alpha_2 = 0$  
(thus proving integrability at three loops). 

The three-loop Hamiltonian $H^{\su(2)}_6$ can be treated as a second-order correction 
to $H^{\su(2)}_2$.  This allows us to numerically evaluate the $O(\lambda^3)$ 
contribution to the spectrum by using second-order Rayleigh-Schr\"odinger
perturbation theory (there is an intermediate state sum involved, but since we
are doing the calculation numerically, this is not a serious problem). There is
also the issue of degeneracy but the existence of a higher conserved charge once
again renders the problem effectively non-degenerate (the details of this argument
will be given in a more detailed study of numerical approaches to the spin chain 
problem \cite{spinchain}). The resulting three-loop data for large-$\alg L$ can be fit
to a power series in $\alg L^{-1}$ to read off the expansion coefficients
$E_{\su(2)}^{3,n}$.  It turns out that, to numerical precision, the coefficients are
non-vanishing only for $n > 5$ (as required by BMN scaling). The first two
non-vanishing coefficients are displayed in table~\ref{NUM_SO(6)_3Loop} for low-lying
levels.  

These results can be compared with string theory predictions derived 
(in the manner described in previous paragraphs) from eqn.~(\ref{exactSO42}), and the
accuracy of the match is displayed in the last column of table~\ref{NUM_SO(6)_3Loop}. 
The important point is that there is substantial disagreement 
with string results at $O(\lambda^3)$
for all energy levels: the low-lying states exhibit a mismatch ranging from roughly 
$18\%$ to $30\%$, and there is no evidence that this can be repaired by taking data 
on a larger range of lattice sizes. There is apparently a general breakdown 
of the correspondence between string theory and gauge theory anomalous dimensions at 
three loops, despite the precise and impressive agreement at first and second order.  
This disagreement was first demonstrated in the two-impurity regime \cite{Callan:2003xr}, 
and additional evidence was presented more recently in the context of a semiclassical
string analysis \cite{Serban:2004jf}. It is therefore perhaps not surprising that
the three-loop disagreement is reproduced in the three-impurity regime, but it provides us with more information which may help to clarify this puzzling phenomenon.
\begin{table}[ht!]
\begin{eqnarray}
\begin{array}{|ccc|}
\hline
 E_{\su(2)}^{(3,7)}/E_{\su(2)}^{(3,6)}  & {\rm String\ Modes\ }(q,r,s) &{\rm \% Error} \\
\hline
16.004		& (1,0,-1) & 33.4\%  \\   
14.114		& (1,1,-2) &   18.8\% \\
14.114		&  (-1,-1,2)&   18.8\%  \\
16.037		&  (2,0,-2)&   33.6\%  \\
14.272		& (1,2,-3) &   21.7\% \\
14.272		&  (-1,-2,3)&   21.7\%  \\
15.561		&  (3,0,-3)&   29.7\% \\
\hline
\end{array} \nonumber
\end{eqnarray}
\caption{Three-loop numerical spectrum of three-impurity 
	$\su(2)$ subsector }
\label{NUM_SO(6)_3Loop}
\end{table}

The same exercise can be repeated for the closed $\su(2|3)$ fermionic sector,
whose string theory dual is comprised of pure fermionic states symmetrized in
$SO(4)\times SO(4)$ indices in either the ${\bf (1,2;1,2)}$ or ${\bf (2,1;2,1)}$
irreps (projected onto $\Pi_\pm$ subspaces).
The spin chain system is embedded in Beisert's $\su(2|3)$ model, where the fermionic 
sector of the Hamiltonian has been recorded up to two-loop order \cite{Beisert:2003ys}.  
Since the relevant points of the numerical gauge/string comparison have already
been made, we will relegate the details of the gauge theory side to \cite{spinchain} 
and simply state the final results.  We also note that Beisert has provided us with the
fermionic part of the \emph{three-loop} vertex in the $\su(2|3)$ sector.  The large-$\alg L$
spectrum of the three-loop contribution will be scrutinized in \cite{spinchain}, but, 
based on the existing evidence at three-loop order, we do not expect an agreement 
with string theory.

In this sector, the $R$-charge and the lattice length are related by $J = \alg L - I/2$.   
The fermionic one- and two-loop string predictions
are therefore found from eqn.~(\ref{exactfermi}) to be
\be
&&E_{\rm Fermi}^{(1,2)} = (q^2 + qr + r^2) \qquad \qquad \kern-2pt
E_{\rm Fermi}^{(1,3)} = 0 \nn\\
&&E_{\rm Fermi}^{(2,4)} = -\frac{1}{4}(q^2 + qr + r^2)^2  \qquad
E_{\rm Fermi}^{(2,5)} = -(q^2 + qr + r^2)^2\ .
\ee
As noted above, this sector does not admit states with equivalent mode indices.

The large-$\alg L$ $\su(2|3)$ fermionic spin chain extrapolations are given at
one-loop order in table~\ref{NUM_fermi_1Loop}.  The two-loop data are obtained using the 
same first-order perturbation theory treatment described above in the 
$\su(2)$ sector; the results are recorded in table~\ref{NUM_fermi_2Loop}.  
The two-loop spectrum is subject to stronger $\alg L^{-1}$ corrections, but the
data are still convincing and could be improved by running the extrapolation
out to larger lattice sizes.  The close agreement for the low-lying levels corroborates 
the match between gauge and string theory up to two-loop order.  
\begin{table}[ht!]
\begin{eqnarray}
\begin{array}{|ccccc|}
\hline
E_{\su(2|3)}^{(1,2)} & E_{\su(2|3)}^{(1,3)} & E_{\su(2|3)}^{(1,3)}/E_{\su(2|3)}^{(1,2)}  
		& {\rm String\ Modes\ }(q,r,s) &  {\rm \%\ Error}  \\
\hline
1+1.3 \times 10^{-10} 	&	-1.9\times 10^{-8}	& -1.9\times 10^{-8} 		&(1,0,-1) & 0.000002\%\\
4-1.0\times 10^{-7} 	&	1.8\times 10^{-5}	& 4.6\times 10^{-6}		&	(2,0,-2)  & 0.0005\%\\
7-2.5\times 10^{-7}  	&	4.4\times 10^{-5}	  & 6.3\times 10^{-6} 		&	(1,2,-3)  & 0.0006\%\\
7-2.5\times 10^{-7}  	&	4.4\times 10^{-5} 	  & 6.3\times 10^{-6}		&	(-1,-2,3)  & 0.0006\%\\
9-3.9\times 10^{-7} 	&	7.9\times 10^{-5}	   & 8.7\times 10^{-6}		&	(3,0,-3)  & 0.0009\%\\
13-4.0\times 10^{-6}  	&	8.2\times 10^{-4}	  & 6.3 \times 10^{-5}		&	(1,3,-4)    & 0.006\%\\
13-4.0\times 10^{-6}  	&	8.2\times 10^{-4}	   & 6.3 \times 10^{-5} 	&	(-1,-3,4)    & 0.006\%\\
16-2.0\times 10^{-5}  	&	4.1\times 10^{-3}	  & 2.6 \times 10^{-4}		&	(4,0,-4)      & 0.003\%\\
19-3.5\times 10^{-5}  	&	7.3\times 10^{-3}	  & 3.8 \times 10^{-4}		&	(2,3,-5)    & 0.004\%\\
19-3.5\times 10^{-5}  	&	7.3\times 10^{-3}	  & 3.8 \times 10^{-4}		&	(-2,-3,5)    & 0.004\%\\
\hline
\end{array} \nonumber
\end{eqnarray}
\caption{Scaling limit of one-loop numerical spectrum of three-impurity 
	$\su(2|3)$ fermionic subsector}
\label{NUM_fermi_1Loop}
\end{table}
\begin{table}[ht!]
\begin{eqnarray}
\begin{array}{|ccccc|}
\hline
E_{\su(2|3)}^{(2,4)} & E_{\su(2|3)}^{(2,5)} & E_{\su(2|3)}^{(2,5)}/E_{\su(2|3)}^{(2,4)}  
		& {\rm String\ Modes\ }(q,r,s) &  {\rm \%\ Error}  \\
\hline
-0.25	  	  &-0.99999 &	  	3.99995  & (1,0,-1) &  0.001\%\\
-4.00006	  &-15.990 & 		3.998  &(2,0,-2) &   0.06\%\\	
-12.251		  &-48.899 &	  	3.992  &(1,2,-3) &   0.2\%\\
-12.251		  &-48.899 &	  	3.992  &(-1,-2,3) &   0.2\%\\
-20.25		  &-80.89 &	  	3.995   & (3,0,-3) &  0.1\%\\
-42.25		  &-168.2 &	  	3.98  &(1,3,-4) &   0.5\%\\
-42.25		  &-168.2 &	  	3.98  &(-1,-3,4) &   0.5\%\\
-64.00		  &-254.6 & 	  	3.98  &(4,0,-4) &   0.6\%\\
-90.26		  &-359.3 &	  	3.98   &(2,3,-5) &   0.5\%\\
-90.26		  &-359.8 &  		3.99  &(-2,-3,5) &  0.3\%\\
\hline
\end{array} \nonumber
\end{eqnarray}
\caption{Scaling limit of two-loop numerical spectrum of three-impurity 
	$\su(2|3)$ fermionic subsector }
\label{NUM_fermi_2Loop}
\end{table}

\clearpage
\section{Conclusions}

The BMN/pp-wave mechanism has emerged as a useful proving ground
for the postulates of the AdS/CFT correspondence.  
When the full Penrose limit is lifted, a rich landscape emerges, 
even in the two-impurity regime, upon which 
the string and gauge theory sides of the duality have exhibited an intricate
and impressive match to two loops in the gauge coupling and first nontrivial
order in the curvature expansion.  
While the conditions under which agreement is obtained
are substantially more demanding in the higher-impurity problem,
we have shown that this agreement is maintained for three-impurity
string states and SYM operators.
We expect that these conclusions will persist for four or more impurities.
Since the Bethe ansatz results of Minahan and Zarembo 
\cite{Minahan:2002ve} provide an all-impurity prediction in the gauge theory,
the methods presented here can easily be employed to perform a simple
check of this statement, for example, in the closed $SO(4)_{S^5}$ sector 
of the string theory for four (or higher) impurity string states.  

Although the two-loop agreement survives at the three-impurity level,
we have also confirmed the previously observed mismatch at three loops
in the gauge theory coupling.  In \cite{Serban:2004jf} it was suggested that
this disagreement may be attributed to an order-of-limits problem.
This notion was made more precise in \cite{Beisert:2004hm}, where
it was noted that a certain class of long-range spin chain interactions
may not survive the small-$\lambda$ expansion in the gauge theory.
A thorough, quantitative understanding of this proposal has yet to be obtained, however.   
In the end, the analyses carried out here will provide an extremely 
stringent test of any proposed solution to this vexing problem.

\section*{Acknowledgements}
We would like to thank John Schwarz and Jonathan Heckman for helpful 
discussions.  We also thank Gleb Arutjunov, Niklas Beisert, Andrei Belitsky,
Charlotte Kristjansen, Didina Serban and Matthias Staudacher for many
helpful comments and suggestions.
This work was supported in part by US Department of Energy 
grants DE-FG02-91ER40671 (Princeton) and DE-FG03-92-ER40701 (Caltech).



\end{document}